 \documentclass[twocolumn]{aastex631}

\shorttitle{Optically thick shocked shell formation}
\shortauthors{Hachisu \& Kato}

\begin{document}

\title{Formation of an optically thick shocked shell in the very fast nova
V1674 Herculis: the origin of superbrightness}


\author[0000-0002-0884-7404]{Izumi Hachisu}
\affil{Department of Earth Science and Astronomy,
College of Arts and Sciences, The University of Tokyo,
3-8-1 Komaba, Meguro-ku, Tokyo 153-8902, Japan}
\email{izumi.hachisu@outlook.jp}

\author[0000-0002-8522-8033]{Mariko Kato}
\affil{Department of Astronomy, Keio University,
Hiyoshi, Kouhoku-ku, Yokohama 223-8521, Japan}




.


\begin{abstract}
V1674 Her is the fastest ($t_2\sim 1$ day) classical nova
in our Galaxy and its absolute $V$ peak of $M_{V,\rm max}\sim -10.2$
is one magnitude brighter than typical very fast novae.  
Such a nova is sometimes called a superbright nova.
Using our fully self-consistent nova outburst model combined 
with the optically thick winds on a $1.35 ~M_\sun$ white dwarf (WD) 
with a mass accretion rate of $1\times 10^{-11} ~M_\sun$ yr$^{-1}$,  
we have clarified that a strong reverse shock arises $0.3$ days after
the outburst, which is just after the maximum expansion of the WD photosphere. 
The shocked shell is optically thick and expanding with the velocity of
$\sim 3500$ km~s$^{-1}$.
Its $V$ brightness reaches maximum of $M_{V,\rm max}=-10.2$
when the shocked shell expands to
$R_{\rm shell}\sim 300 ~R_\sun$ on day $\sim 0.7$. 
After that, the shocked shell turns to optically thin and becomes fainter
than the brightness of free-free emission from the nova wind. 
In chronological order, the optical brightness of free-free emission 
reaches maximum of $M_V=-9$ on day 0.3.
However, it is overtaken on day 0.5--0.7 by the
$\sim$1 mag brighter luminosity of the optically thick shocked shell. 
The GeV gamma-ray flux reaches maximum on day 0.4 because
the gamma-rays are emitted by the shock that arises on day 0.3.
Our model consistently explains both the superbrightness and chronological 
order that the gamma-ray peak precedes substantially before the optical
$V$ peak.  We also present a similar light curve model
for another superbright nova V1500 Cyg.
\end{abstract}





\section{Introduction}
\label{introduction}

\begin{figure*}
\epsscale{0.9}
\plotone{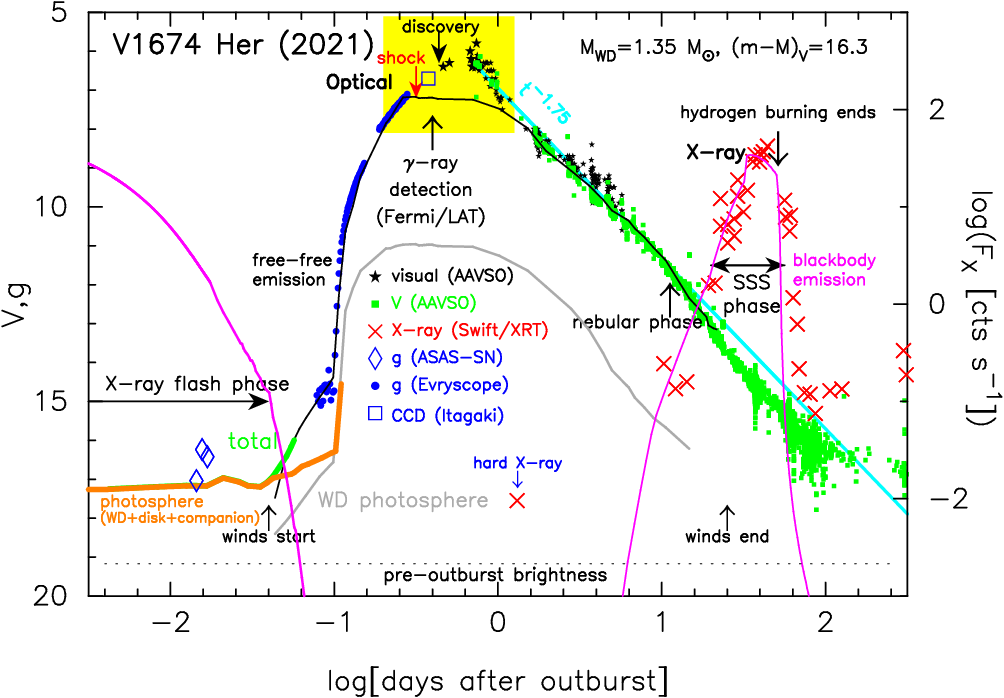}
\caption{
Summary of the visual, $V$, $g$, and X-ray (0.3--10.0 keV) light curves
of V1674 Her for both models and observations. The discovery date is
indicated by the downward black arrow labeled ``discovery.''
The $V$ and visual data are taken from the archive of 
the American Association of Variable Star Observers (AAVSO).
The All-Sky Automated Survey for Supernovae (ASAS-SN) $g$, 
Evryscope $g$, and Itagaki's unfiltered CCD data are from \citet{qui24}.
The X-ray count rates are from the Swift website \citep{eva09}. 
We add theoretical $V$ (black line) and X-ray (magenta line) light curves
based on \citet{kat25hs}'s fully self-consistent nova outburst model.
We set our theoretical outburst day ($t=0$ at epoch B in their Figure 1(a))
to be HJD 2,459,377.68($=$UT 2021 June 12.18).  The WD model has the mass of
$M_{\rm WD}= 1.35 ~M_\sun$ with the mass-accretion rate of
$\dot{M}_{\rm acc}= 1\times 10^{-11} ~M_\sun$ yr$^{-1}$.
The model $V$ light curve (black line) is calculated from free-free emission
from nova winds \citep{hac25kv1674her2} whereas the model X-ray light curve
(magenta line) is calculated from the blackbody emission
from the WD photosphere (0.3--10.0 keV).
The thick orange line shows the photospheric $V$ light curve of the WD,
accretion disk, and companion star and the light gray line corresponds
only to the WD photosphere, which are taken from \citet{hac25kv1674her2}.
The straight thick cyan line labeled
$t^{-1.75}$ denotes the universal decline law of $L_V\propto t^{-1.75}$
\citep{hac06kb}, where $L_V$ is the $V$ band luminosity.  
The $V$ band distance modulus $\mu_V\equiv (m-M)_V= 16.3$, the distance
$d=8.9$ kpc, and the extinction $E(B-V)=0.5$ toward V1674 Her are taken
from \citet{kat25hs}.  There is a gap between the theoretical free-free
emission model light curve (black line) and the observation,
as demonstrated in the yellow-shadowed area.
We also show the pre-outburst brightness of $g = 19.17$ \citep[dotted 
line; ][]{qui24} 1.7 days before the nova outburst ($t=-1.7$ day).
See the main text for more detail.
\label{v1674_her_v_x_observation_only_logscale}}
\end{figure*}

The classical nova V1674 Her (Nova Herculis 2021) was discovered
at 8.4 mag on UT 2021 June 12.537 by Seiji Ueda (cf. CBET 4976).
It has been observed in multiple wavelengths, from radio, NIR,
optical, UV, and X-ray, to gamma-ray
\citep{dra21, woo21, lin22, pat22, ori22, sok23, bha24, hab24, qui24}.
One of the remarkable features of V1674 Her is rich observational data in
the very early phase of the outburst, that is, in the pre-discovery period
as shown in Figure \ref{v1674_her_v_x_observation_only_logscale}.
Unfortunately no X-ray flash was observed, but
dense optical data toward maximum were obtained over 10 magnitudes rise 
\citep{qui24}. 
Such a dense time series in the very early phase
is the first obtained in classical novae. 
This enable us to fit our model light curves with the observation 
during the full period of the nova outburst, i.e., from  
the extremely early phase to the very late phase of the 
nova outburst \citep{kat25hs, hac25kv1674her2}. 
Figure \ref{v1674_her_v_x_observation_only_logscale} also shows the
model light curve (black line) that is well fitted with the V1674 Her
observation \citep{kat25hs}.


\begin{figure*}
\epsscale{1.0}
\plottwo{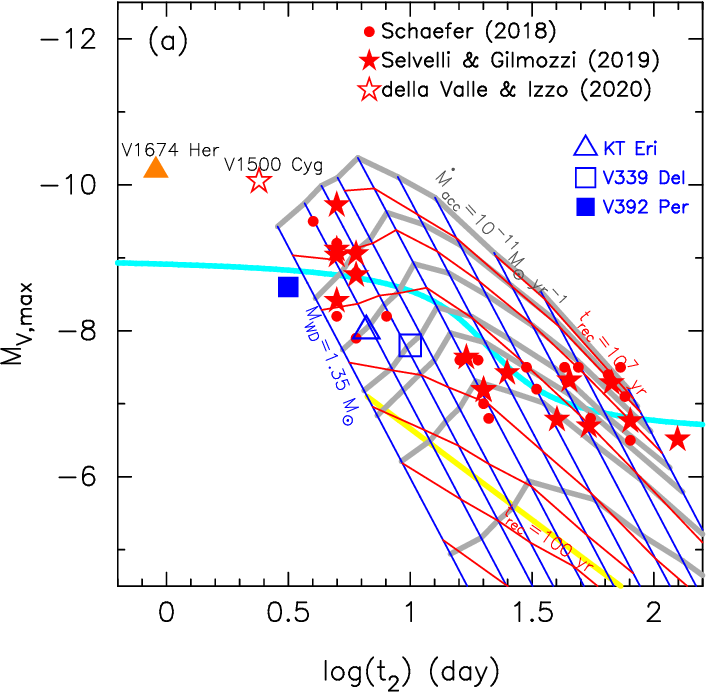}{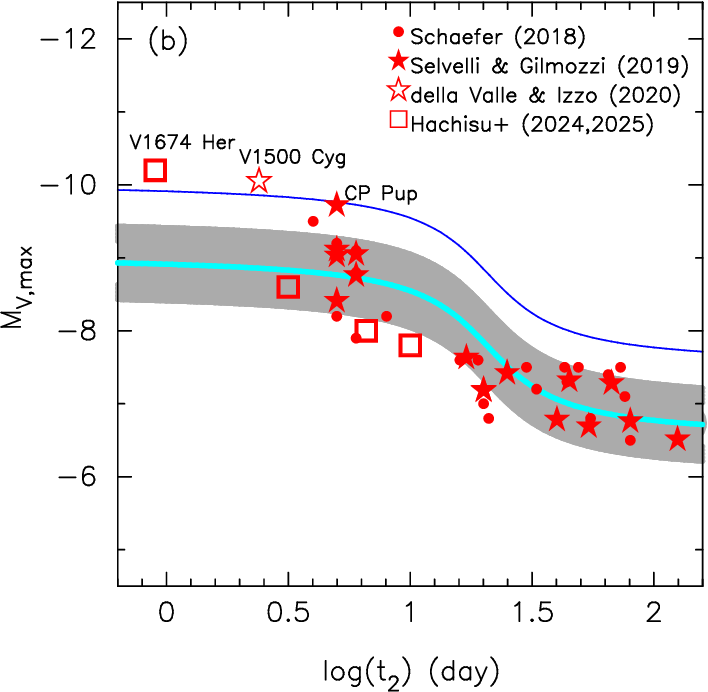}
\caption{
Maximum $V$ magnitude versus rate of decline (MMRD) diagram,
$\log (t_2)$-$M_{V,\rm max}$, for classical novae.
(a) The blue lines indicate theoretical model equi-WD mass lines, from left
to right, 1.35, 1.3, 1.25, 1.2, 1.1, 1.0, 0.9, 0.8, 0.7, and $0.6~M_\sun$;   
the thick solid gray lines denote model equi-mass accretion rate 
($\dot M_{\rm acc}$) lines, from lower to upper, $3\times 10^{-8}$,
$1\times 10^{-8}$, $5\times 10^{-9}$, $3\times 10^{-9}$, $1\times 10^{-9}$, 
$1\times 10^{-10}$, and $1\times 10^{-11} M_\sun$~yr$^{-1}$;
the red lines represent model equi-recurrence time lines, from lower to upper,
$t_{\rm rec}= 30$, 100, 300, 1000, 10000, $10^5$, $10^6$, and $10^7$~yr.
These lines are taken from \citet{hac20skhs} based on the optically thick
nova wind model \citep{kat94h}
and nuclear runaway model calculation of mass accretion onto each WD. 
The brightnesses of novae are calculated from free-free emission
luminosity of Equation (\ref{free-free_flux_v-band}).
The thick yellow line corresponds to the $x_0\equiv M_{\rm env}/ M_{\rm sc}
=2$ line, where $M_{\rm env}$ is the hydrogen-rich envelope mass at the
optical maximum and $M_{\rm sc}$ the scaling mass.
Their assumed scaling law for $\dot{M}_{\rm wind}$
and $M_{\rm env}$ is valid only for $x_0\gtrsim 2$.  Therefore, below 
the yellow line ($x_0<2$), the brightnesses of these models are
not accurate \citep[see ][for details]{hac20skhs}.
We overplot filled red circles taken from ``Golden sample'' of
\citet{schaefer18}, filled stars from \citet{sel19},
and open star (V1500~Cyg) from \citet{del20i}.
The three novae (KT Eri, V339 Del, and V392 Per) are
taken from \citet{hac25kw}, \citet{hac24km}, and \citet{hac25kv392per},
respectively. 
The thick solid cyan line indicates the empirical
line for the MMRD relation obtained by \citet{del20i}.
The two novae, V1674 Her (orange triangle) and V1500 Cyg (unfilled
red star), are located outside the region of \citet{hac20skhs}.
The peak brightnesses of these two novae cannot be reproduced
by the free-free emission model light curves, which indicates
that the energy source is different from free-free emission.  See Sections
\ref{evolution_shocked_shell} and \ref{light_curve_v1500_cyg}, respectively,
for their reasons. 
(b) Same as panel (a), but we show only the position of each nova and
empirical MMRD line of \citet{del20i}.   The thick cyan line
indicates the same as the thick cyan line in panel (a), and light-gray
shadow line corresponds to its $\pm 0.5$ mag region.
The blue line is 1 mag above the thick cyan line. 
\label{vmax_t2_selvelli2019_schaefer2018_2fig}}
\end{figure*}

\subsection{Optical rise with no shock powering}
\label{optical_rise_intro}

\citet{kat25hs} calculated a light curve model of V1674 Her
based on a fully self-consistent nova explosion model of a 
$1.35 ~M_\sun$ white dwarf (WD) with a mass-accretion rate to the WD of
$1\times 10^{-11} ~M_\sun$ yr$^{-1}$.
Their free-free emission model $V$ light curve (the black line in
Figure \ref{v1674_her_v_x_observation_only_logscale}) reproduces well
the observed $V$ and $g$ light curves of V1674 Her, including the very fast
rising phase and the decay phase after optical maximum.
Modeling a detailed $V$ light curve of the very early rising phase
of V1674 Her, \citet{hac25kv1674her2} showed that the earliest $g=$16--17
mag detections \citep{qui24} corresponds to the X-ray flash phase
(the observed three open blue diamonds and model thick orange line
in Figure \ref{v1674_her_v_x_observation_only_logscale}).  
This is the first indirect optical detection of an X-ray flash phase
of a nova, even though there is no X-ray observation during the X-ray
flash phase \citep{hac25kv1674her2}.

\citet{hac25kv1674her2} further clarified that the rapid increase in
the early optical brightness is caused by the envelope structure change
due to a large variation of radiative opacity in the envelope.
Thus, this rapid increase confirmed that the driving force of nova
envelope expansion and/or nova wind is radiative-pressure gradient
owing to the radiative opacity in the envelope.  In other words, we do not
need shock powering to explain the rapid rising phase of the nova.  
Reproducing well the light curves, from the X-ray flash phase to
the supersoft X-ray source (SSS) phase, confirmed that the entire 
nova evolution is governed by expansion and the ensuing optically thick wind
mass loss of the WD envelope.

\subsection{Formation of a shock}
\label{shock_formation_intro}

Hard X-ray and GeV gamma-ray emissions have been often
observed in classical novae.
Hard X-rays were detected in an intermediate phase of a nova outburst
\citep[e.g.,][]{llo92ob, bal98ko, muk01i}.
GeV gamma-ray emissions were observed in an early phase of a nova,
just from the post-maximum phase, and continues a few tens of days
\citep[e.g.,][]{abd10, ack14aa, li17mc, gor21ap}.

These high-energy (hard X-rays and GeV $\gamma$-rays)
emissions are considered to originate from strong shocks
between shells ejected with different velocities
\citep{cho14ly, met15fv, mar18dj}.
If the inner shell (later ejected) has a larger velocity than that of
the outer shell (earlier ejected), the inner one can catch up with the
outer one and forms a shock wave \citep[e.g.,][]{muk19s, ayd20ci, ayd20sc}.
The assumption of multiple shell ejection is the key of this idea.

Such a multiple shell ejection was suggested
from both optical and high-energy emissions from novae.
There is, however, no theoretical explanation has been presented that
naturally explains all these different wavelength observations
based on nova explosion models \citep[see][for a recent review]{cho21ms}.

From the theoretical point of view, many numerical calculations
have been presented from the early thermonuclear runaway to the
extended phase of nova outbursts \citep[e.g.,][]{
pri92k, pri95k, epels07, sta09ih, den13hb, chen19, kat22sha, kat22shb}.
These works clarified that mass ejection is continuous,
no shock arises at the thermonuclear runaway, 
and no multiple distinct mass ejection occurs.

\citet{hac22k} showed that a shock arises outside the WD 
photosphere based on \citet{kat22sha}'s
fully self-consistent nova explosion model.  
\citet{hac22k} found that a strong reverse shock inevitably arises far outside
the WD photosphere after the maximum expansion of the WD photosphere.
This is because the velocity of nova ejecta continuously and smoothly
increases with time after the maximum expansion
\citep[see, e.g., Figure 1 of ][]{hac22k}.
The later ejected matter has a larger 
expansion velocity so that it catches up with the former ejected matter  
and makes a strong shock. Thus, a shock is formed after the maximum expansion
of the WD photosphere ($=$optical maximum) 
and propagates far outside the WD photosphere. 
This shock formation mechanism reasonably explains gamma-ray emission 
and hard X-ray detection/nondetection in classical novae \citep[e.g.,
YZ Ret, V339 Del, and V392 Per in][respectively]{hac23k, hac24km,
hac25kv392per}. 

Novae sometimes accompany GeV gamma-rays and its origin
is related to strong shocks \citep[e.g.,][for a recent review]{cho21ms}.
In V1674 Her, GeV gamma-rays were also detected \citep{sok23}. 
\citet{hac22k}'s interpretation of gamma-ray emission is also 
based on the shock origin. Their theory predicts that gamma-ray emission 
should be detected only after the maximum expansion of the WD photosphere
($=$optical peak).
In V1674 Her, however, the GeV gamma-ray flux peaked on day $\sim$0.4
\citep{sok23} substantially before the optical maximum on day $\sim$0.8
\citep{hab24}, as shown by the yellow-shaded region
in Figure \ref{v1674_her_v_x_observation_only_logscale}. 
This chronological order that the gamma-ray peak substantially
precedes the optical maximum is not consistent with Hachisu \& Kato's 
shock formation theory, which suggests that the optical maximum is first
and then the gamma-ray peak, mentioned above.

\subsection{Superbright novae}
\label{superbright_novae_intro}


V1674 Her is an extremely bright nova.
\citet{kat25hs}'s free-free emission model light curve (black line)
is $\sim$1 mag fainter than the observed $V$ peak, as shown in Figure 
\ref{v1674_her_v_x_observation_only_logscale}.
The absolute $V$ magnitude of V1674 Her is 
$M_{V,\rm max}= -10.2$ from the $V$ band distance modulus
of $\mu_V\equiv (m-M)_V=16.3$, i.e., the distance of $d=8.9$ kpc
and the extinction of $E(B-V)=0.5$ \citep{kat25hs}.
The 2 mag decay time from the $V$ maximum, $t_2=0.904$ days,
was given by \citet{hab24}.  We plot this maximum magnitude versus
rate of decline (MMRD) point on the $t_2$-$M_{V,\rm max}$ diagram, i.e., in
Figure \ref{vmax_t2_selvelli2019_schaefer2018_2fig}.  Note that 
different $t_2$ times are reported to be 1.1--1.2 days \citep[e.g.,][]{sok23},
which are not plotted in Figure \ref{vmax_t2_selvelli2019_schaefer2018_2fig}
but they are located closely to each other.
We do not use a specific $t_2$ time in our modeling.

In Figure
\ref{vmax_t2_selvelli2019_schaefer2018_2fig}(a),
we add other nova positions taken from several literatures as well as
our theoretical results (various lines) that are calculated
from model free-free light curves of novae in \citet{hac20skhs}.
The position (filled orange triangle) of V1674 Her is far outside
the region of our model free-free light curves. 

In Figure \ref{vmax_t2_selvelli2019_schaefer2018_2fig}(b), the peak $V$
magnitude of V1674 Her is $\sim$1 mag brighter than the typical
MMRD relation (thick cyan line) in the $t_2$-$M_{V,\rm max}$ diagram.
Many novae are located around ($\pm 0.5$ mag) this line, although a few
novae are $\gtrsim 1$ mag (blue line) brighter than the thick cyan line.
\citet{del91} dubbed them ``super-bright novae'' that reached the absolute
$V$ brightness of $M_{V, \rm max} \lesssim -10$ mag and $\gtrsim$1 mag
brighter than the typical MMRD line \citep[thick cyan line,][]{del20i}.
The classical nova V1500 Cyg is a prototype superbright nova defined by
\citet{del91} in our Galaxy.  
The origin of these brightest novae has not been clarified yet.

\subsection{Objectives}
\label{objectives_intro}

The aim of this paper is to solve the two problems,
(1) the superbright luminosity at the optical peak and
(2) the inverse chronological order of the emergence of 
gamma-ray emission before the optical peak,  
that \citet{kat25hs} had not explained.
Here, we propose an idea that, if 
the expanding shocked shell is optically thick, its photospheric 
brightness could exceed the flux of free-free emission
from the nova winds,  
and also the shock luminosity could delay the optical maximum
until after the emergence of gamma-ray emission.

This paper is organized as follows. First we present quick physical
interpretation of observation based on our $1.35~M_\sun$ WD model
in Section \ref{quick_interpretation}.
We construct our model light curves based on our optically thick 
shocked shell model and compare with the observation of V1674 Her
in Section \ref{evolution_shocked_shell}.  We also show our model $V$
light curve calculation for another superbright nova V1500 Cyg in
our Galaxy in Section \ref{light_curve_v1500_cyg}. 
Discussion and conclusions follow in Sections \ref{sec_discussion}
and \ref{sec_conclusion}, respectively.


\begin{figure*}
\gridline{\fig{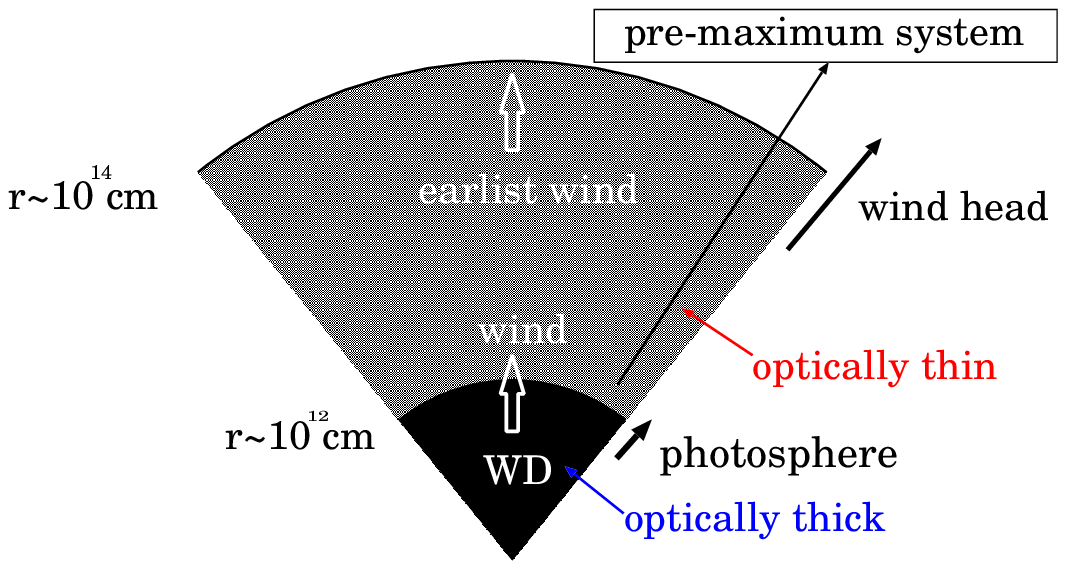}{0.22\textwidth}{(a) pre-maximum expansion of nova
(WD) photosphere}
          \fig{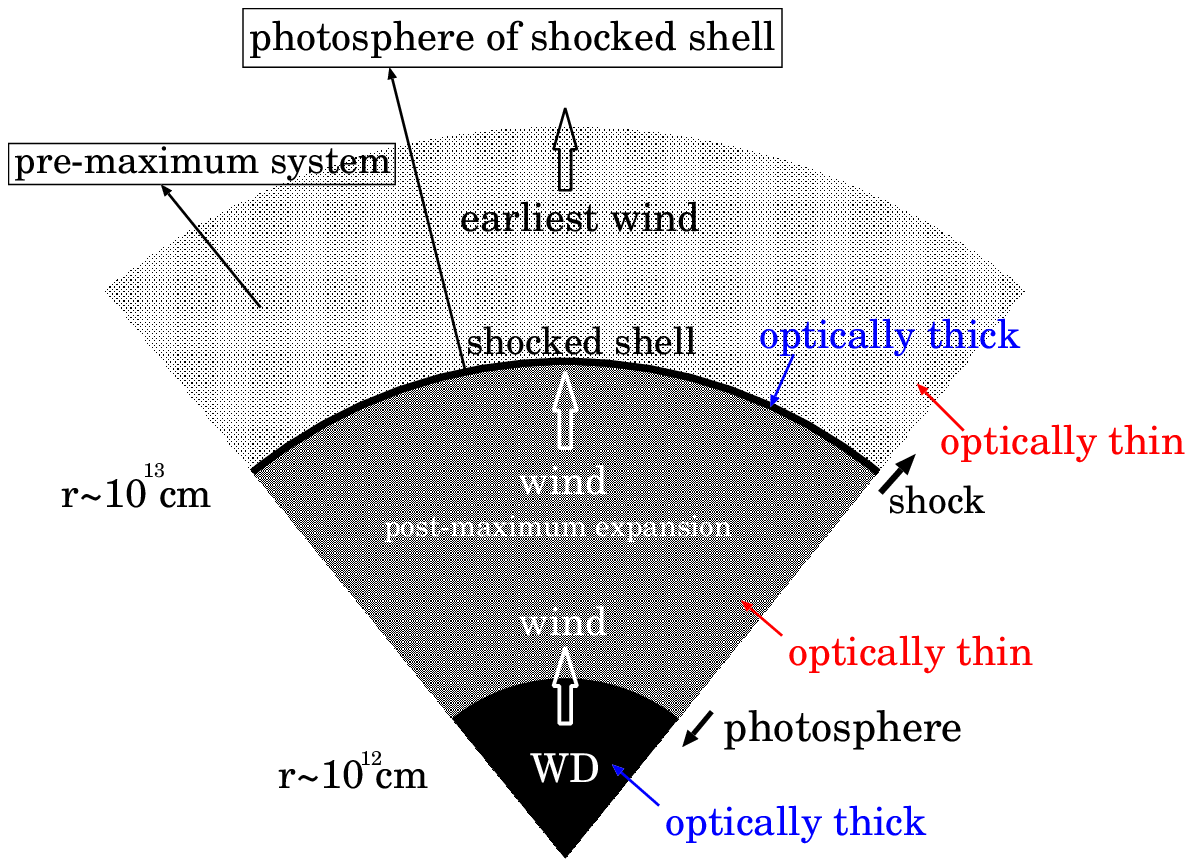}{0.35\textwidth}{(b) post-maximum expansion of nova
(WD) photosphere}
          \fig{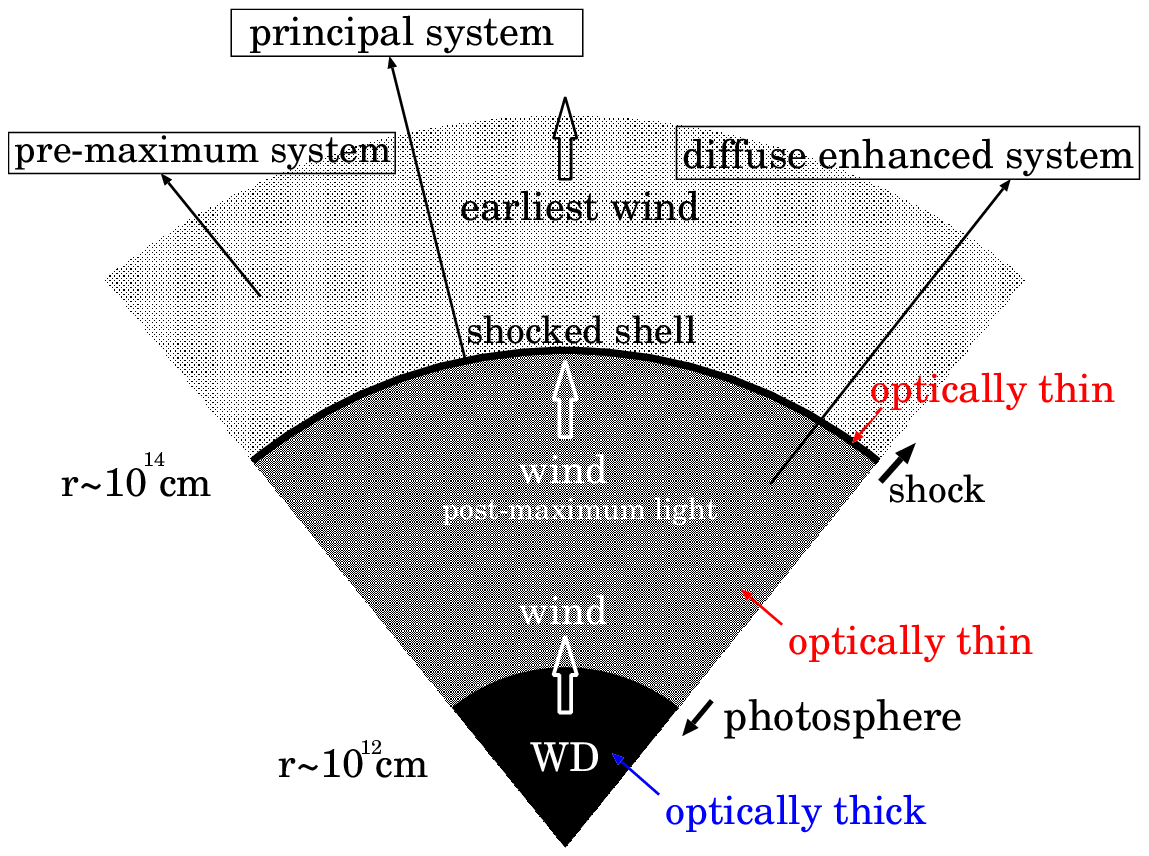}{0.4\textwidth}{(c) post-maximum light}
          }
\gridline{
          \fig{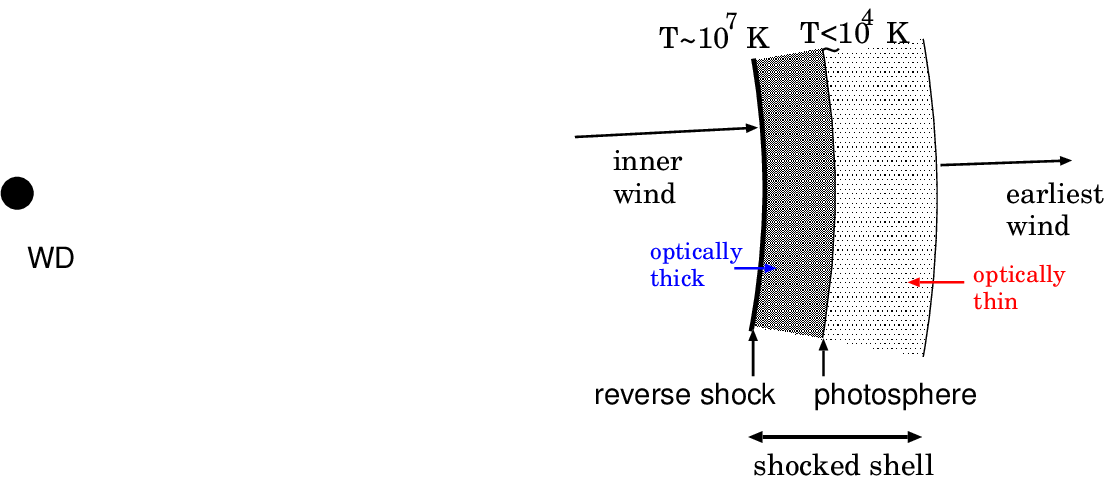}{0.75\textwidth}{(d) configuration in the shocked shell}
          }
\caption{
Cartoon for our V1674 Her nova model in the early phase.  
(a) The nova (WD) photosphere expands over $R_{\rm ph} \sim 0.1 ~R_\sun$ and 
optically thick winds are accelerated deep inside the photosphere
\citep{kat22sha,kat25hs}.  The wind itself becomes optically thin
outside the photosphere.  The nova (WD) photosphere is further expanding.
The earliest wind forms the pre-maximum absorption/emission line system
\citep{mcl42} outside the WD photosphere ($r > R_{\rm ph}$).
(b) After maximum expansion of the nova (WD) photosphere, the photosphere
is receding. A strong shock arises outside the WD photosphere \citep{hac22k}.
The shocked shell is so dense that the optical depth $\tau_{\rm shell}$
of the shell is larger than unity (optically thick)
just after the shock arises.  The shocked shell emits gamma-rays.
(c) The shocked shell is further expanding and its optical depth
$\tau_{\rm shell}$ gradually decreases to less than unity (optically thin).
The shocked shell is geometrically thin and optically thin.
The whole ejecta is divided into three parts, outermost expanding gas
(earliest wind), shocked shell, and inner wind.
These three parts contribute to pre-maximum, principal,
and diffuse enhanced absorption/emission line systems \citep{mcl42},
respectively, as proposed by \citet{hac22k, hac23k}.
The velocity of principal system is typically about a half of that of
diffuse enhanced system \citep{mcl42, hac22k}.
The optically thin shocked shell emits thermal hard X-rays.
(d) An enlargement of the shocked layer in panel (b).  We plot
locations of the reverse shock, hydrogen recombination front (photosphere),
outermost edge of the shocked shell (optically thin layer).
The photosphere of the shocked shell emits photons like a supergiant.
\label{wind_shock_config}}
\end{figure*}

\section{Quick interpretation of observation}
\label{quick_interpretation}

Figure \ref{v1674_her_v_x_observation_only_logscale} summarizes the 
visual, $V$, $g$, and X-ray (0.3--10.0 keV) light curves of V1674 Her. 
The source of observational data are described in the caption.
It also shows the model $V$ and X-ray light curves 
for a $M_{\rm WD}= 1.35 ~M_\sun$ WD with a mass-accretion rate to the WD
of $\dot{M}_{\rm acc}= 1\times 10^{-11} ~M_\sun$ yr$^{-1}$ taken from
\citet{kat25hs} and \citet{hac25kv1674her2}.

The model $V$ light curve (black line) is calculated from free-free emission
\citep[Equation (3) of ][or Equation(\ref{free-free_flux_v-band})]{kat25hs}
of nova winds whereas the model X-ray light curve (magenta line) is calculated
from the blackbody emission of the WD photosphere (0.3--10.0 keV).
Our free-free emission model light curve (black line) 
reasonably reproduces the $V$, visual, and $g$ observations
except for during around the optical peak (yellow-shaded region)
and very early phase of ASAS-SN $g$ data (three open blue diamonds).
We adopt our outburst day ($t=0$) of 
$t_{\rm OB}=$HJD 2,459,377.68 ($=$UT 2021 June 12.18) after \citet{kat25hs}.

The thick orange line shows the summation of the photospheric $V$ light
curves of the WD, accretion disk, and companion star while the light-gray
line corresponds only to the WD photosphere, the data of which are taken
from Figure 6 of \citet{hac25kv1674her2}.
Here, we adopt the companion mass of $0.26 ~M_\sun$
after \citet{qui24} and the orbital period of 3.67 hr (0.1529 days)
and its ephemeris after \citet{pat22}.  
The disk size is 0.85 times the effective Roche lobe
radius and the thickness of the disk is 0.05 times the disk radius 
\citep[see][for detail]{hac25kv1674her2}.
The inclination angle of the binary is assumed to be $67\arcdeg$ \citep{hab24}.
We also show the quiescent brightness of $g = 19.17$ \citep[dotted 
line; ][]{qui24} 1.7 days before the nova outburst ($t=-1.7$ days).

\subsection{X-ray flash phase (0--0.04 days)}
\label{x-ray_flash_phase}

\citet{kat25hs} calculated a full cycle of a nova outburst with
a Henyey type evolution code combining optically thick nova winds.
 After hydrogen ignites to trigger an explosion, the photospheric
temperature of the WD rises up to $k T_{\rm ph}\sim 86$ eV and then
turns to decrease, where $k$ is the Boltzmann constant and
$T_{\rm ph}$ the photospheric temperature of the WD.
Thus, the WD photosphere emits dominantly supersoft X-ray photons.  
The rising phase of the X-ray flash was depicted in Figure 1 (HR 
diagram) of \citet{kat25hs} or Figure 5 (X-ray light curve) of
\citet{hac25kv1674her2}. 
When the wind mass loss starts, the X-ray flash phase ends.

In this X-ray flash phase, the WD photospheric luminosity had quickly
increased to near the Eddington limit 0.0012 days after the outburst.
For the optical $V$ band, the WD photosphere does not
contribute at all, and instead the irradiated disk and companion star 
photospheres become so bright and contribute to the optical $V$ band 
\citep[thick orange line: see][for detail]{hac25kv1674her2}.
 
Unfortunately the X-ray flash itself was not observed in X-ray,
but the three ASAS-SN $g$ band observations (open blue diamonds in Figure
\ref{v1674_her_v_x_observation_only_logscale}) are the first optical
detection of an X-ray flash phase \citep{hac25kv1674her2}.   
It should be noted that no shocks arise inside the WD photosphere 
\citep[see ][for details]{hac22k}.

\subsection{Optically thick wind phase (0.04--27 days)}
\label{wind_phase}

When the WD photosphere expands to $R_{\rm ph}\sim 0.1 ~R_\sun$
and its temperature decreases to $T_{\rm ph}\sim 150,000$ K,
optically thick winds emerge from the WD photosphere ($t=0.04$ days).
The X-ray flux quickly decays because of the quick decrease in the
photospheric temperature.  We identify the end of the X-ray flash phase
on day 0.04 after \citet{kat25hs} and \citet{hac25kv1674her2}.

In the wind phase, the $V$ band luminosity is dominated by free-free
emission from the optically thin ejecta outside the WD photosphere.
It should be noted that optically thick winds are
accelerated deep inside the WD photosphere but the wind itself becomes
optically thin outside the WD photosphere
as illustrated in Figure \ref{wind_shock_config}(a).

\citet{kat25hs} calculated the free-free $V$ luminosity as
\begin{equation}
L_{V, \rm ff,wind} = A_{\rm ff} ~{{\dot M^2_{\rm wind}}
\over{v^2_{\rm ph} R_{\rm ph}}}
\label{free-free_flux_v-band}
\end{equation}
\citep{hac06kb,hac20skhs}.
Here, $\dot{M}_{\rm wind}$ is the wind mass loss rate, 
$v_{\rm ph}$ is the velocity at the photosphere, 
and $R_{\rm ph}$ is the photospheric radius. 
See Equation (3) in \citet{kat25hs} for details on the coefficient
$A_{\rm ff}$ and how to determine it for V1674 Her.

\begin{figure*}
\epsscale{0.75}
\plotone{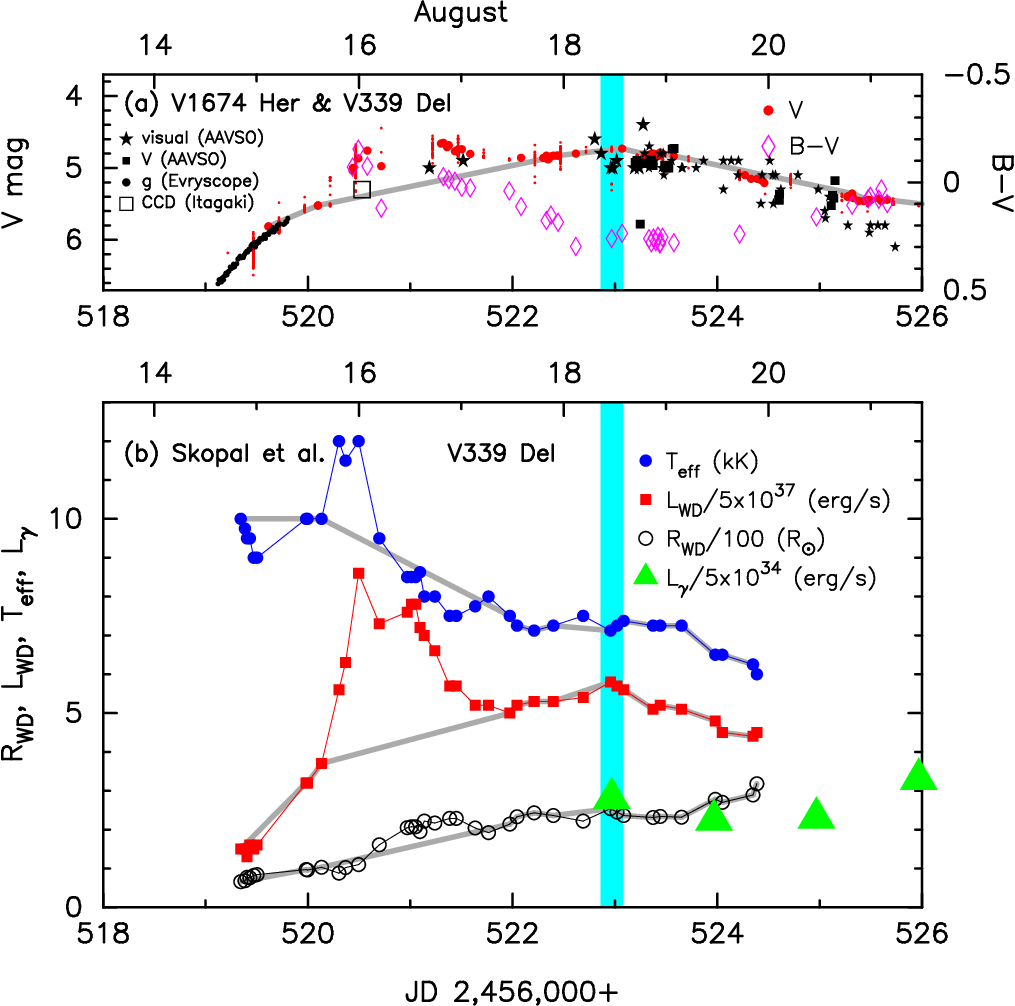}
\caption{(a) The early $V$, visual, and $g$ light curve of V1674 Her
(black symbols) and V339 Del (small and large filled red circles).
The data on V339 Del are taken from Figure 2 of \citet{hac24km}.
The timescale of V1674 Her is expanded by 7.55 and its $V$ magnitude
is shifted up by $\Delta V= 1.4$.  The $V$, visual, and $g$ data of
V1674 Her are the same as those in Figure 
\ref{v1674_her_v_x_observation_only_logscale}
and correspond to the yellow-shaded region in
Figure \ref{v1674_her_v_x_observation_only_logscale}.
We add $B-V$ color evolution of V339 Del (open magenta diamonds).
The large symbols of V339 Del denote the data taken from \citet{mun15mm},
\citet{bur15a}, SMARTS \citep{wal12bt}, and OKU \citep{hac24km},
while the small filled red circles are taken from AAVSO.
The broad gray line indicates our approximations of the $V$ magnitude.
The vertical broad cyan line denotes the epoch of the global optical peaks
of V339 Del, i.e., Aug. 18.47$\pm$0.11 \citep{sko14dt}.
(b) The temporal developments of the effective temperature $T_{\rm eff}$,
luminosity $L_{\rm WD}$, and radius $R_{\rm WD}$ of the pseudo-photosphere,
taken from \citet{sko14dt}.  The broad gray lines indicate our
approximations to the temporal developments of each value
when we exclude the early
fluctuations in the data of V339 Del. The luminosity and radius depend on
the assumed distance to the nova.  \citet{sko14dt} assumed $d=3$~kpc,
so that the luminosity and radius should be translated from the 
original values to the true values according to
$L_{\rm WD} \propto (d/3{\rm ~kpc})^2$ and
$R_{\rm WD} \propto (d/3{\rm ~kpc})$, respectively.
The three thin blue, red, black lines connect each data.
We also add GeV gamma-ray fluxes \citep[filled green triangles;][]{ack14aa}.
\citet{ack14aa} assumed $d=4.2$ kpc, so the
gamma-ray luminosity depends on $L_{\gamma} \propto (d/4.2{\rm ~kpc})^2$.
\citet{hac24km} determined the distance of V339 Del to be 2.1 kpc.
}\label{v1674_her_v339_del_v_skopal_wd_photo}
\end{figure*}

The free-free emission luminosity depends strongly on the wind mass-loss
rate of $\dot{M}_{\rm wind}$ as shown in Equation 
(\ref{free-free_flux_v-band}).  The wind mass-loss rate abruptly
increases on day 0.1 because 
the photospheric temperature of the WD
decreases to $\log T_{\rm ph}~($K$) < 5.2$ 
and the continuum-radiation pressure increases inside the envelop 
that accelerate the winds \citep{hac25kv1674her2}.

In Figure \ref{v1674_her_v_x_observation_only_logscale},
after the optical $V$ peak on day 0.7--0.8,
the light curve decays almost along with
the line of the universal decline law, $L_V \propto t^{-1.75}$
(thick cyan line), where $L_V$ is the $V$ band luminosity and
$t$ is the time from the outburst, as many classical novae do 
\citep[e.g., ][]{hac06kb, hac15k, hac16k, hac19kb}.
\citet{hac06kb} calculated many light curves for various WD masses
and chemical compositions based on Equation (\ref{free-free_flux_v-band}),
and found that their decline slopes are close to the slope of 
$L_V \propto t^{-1.75}$. They dubbed these light curve slopes
``the universal decline law.''
After the nova entered the nebular phase, its brightness drops more rapidly
than the line of $L_V \propto t^{-1.75}$.
We can interpret this trend with the quicker
decrease in the wind mass-loss rate after day 11.

\subsection{Shock formation (0.32--1.2 days)}
\label{shock_formation}

After the optical maximum of free-free emission (black line in
Figure \ref{v1674_her_v_x_observation_only_logscale}),
a strong shock arises and then propagates
toward far outside the WD photosphere \citep{hac22k, hac23k}. 
The velocity at the photosphere $v_{\rm ph}$ decreases with time
before the maximum expansion of the WD photosphere, but turns
to increase after that.  In the post-maximum expansion phase (Figure
\ref{wind_shock_config}(b)), the wind ejected later is catching up
the matter previously ejected, which causes a strong shock far
outside the WD photosphere \citep{hac22k}.

If the shocked shell is optically thin, the nova magnitude can be described
by the free-free emission (black line in Figure 
\ref{v1674_her_v_x_observation_only_logscale}).  If the shocked shell is
optically thick, the free-free emission from the inner wind 
(inside the shocked shell) is absorbed by the shocked shell.
As a result, we observe only the emission from the shocked shell.  
\citet{kat25hs} did not examine whether the shocked shell is optically
thick or not.  In the next section, we will examine the properties
of the shocked shell and obtain the light curve including a contribution from
this shocked shell photosphere that emits photons like a supergiant
(Figure \ref{wind_shock_config}(d)).

\section{Evolution of the shocked shell}
\label{evolution_shocked_shell}

\subsection{Photospheric temperature evolution of
optically thick shocked shell}
\label{photospheric_temperature_shocked_shell}

Figure \ref{wind_shock_config}(a)-(c) illustrates how the strong shock
arises and propagates outward in the ejecta of V1674 Her
\citep[see][for more details on the shock propagation]{hac22k}.
The optical depth $\tau_{\rm shell}$ of the shocked shell will be estimated
below in Section \ref{optical_depth_shock}; it is larger than unity
(optically thick, $\tau_{\rm shell}\gtrsim1.0$) from day 0.32
(the emergence epoch of a shock) until the optical peak on day 0.6--0.7,
followed by a gradual decrease down to $\tau_{\rm shell}\sim0.3$ on day 1.2.  
Figure \ref{wind_shock_config}(d) illustrates a close-up view of the 
shell structure when the shocked shell is optically thick in V1674 Her. 

Its photosphere would correspond to the recombination front
of hydrogen as seen in shock-heated expanding ejecta of
Type IIP supernovae (SNe IIP) \citep[e.g,][]{dub25}. 
Thus, the temperature around the recombination front (photosphere)
may be about 10,000 K or less.

The luminosity of the shocked shell photosphere can be approximately 
calculated as a supergiant spectrum with the effective temperature of
$T_{\rm eff}$.  \citet{mun21vd} obtained $B-V=0.719$ on day 0.72 for
V1674 Her.  We estimate $T_{\rm eff}=7800$ K from the intrinsic color
of $(B-V)_0=0.719 - 0.55=0.169$ with $E(B-V)=0.55$ given by \citet{mun21vd}.


However, we have no information on the color/temperature evolution
around the optical peak.  We instead adopted a temperature evolution
which mimics the temperature evolution of the classical nova V339 Del
in which gamma-rays are also detected. 

Figure \ref{v1674_her_v339_del_v_skopal_wd_photo}(a) compares 
the close-up view of the light curves of V1674 Her and V339 Del around
the optical peak.  If we expand the timescale of V1674 Her by 7.55,
the two light curves well overlap.

Many novae show very similar spectra near their optical maxima, that is,
those of F supergiants, regardless of the speed class.
Therefore, \citet{van87} suggested a common color, 
$(B-V)_0= +0.23 \pm 0.06$, at optical maximum of a nova.
This also means that their color evolutions are similar near
their optical maxima even though their optical $V$ magnitudes $M_V$ are
different.  The color temperature of a nova near optical maximum
is closely related to the photospheric temperature of an F supergiant.

Our fully self-consistent nova explosion code calculates only inside of
the WD photosphere \citep{kat25hs}, and does not calculate the formation
of a shock that occurs outside the WD photosphere.  
Here, we adopt a simplified photospheric model for a shocked shell
instead of radiation hydrodynamic calculation on the ejecta
outside the WD photosphere.

Figure \ref{v1674_her_v339_del_v_skopal_wd_photo}(b) shows 
the evolutions of temperature, luminosity, and radius of V339 Del taken from 
\citet{sko14dt} in the optically thick photosphere phase, or
in the fireball phase by their terminology.
To avoid confusion with freely extended definitions of ``fireball,''
we repeat \citet{gehrz88}'s original definition:\\  \noindent
``$<$Fireball Expansion$>$\\ \noindent
Photometry of novae at outburst shows that the ejecta radiate like hot
($T =$ 6000--10,000 K) blackbodies that are expanding with time (63, 82,
95, 98, 168). Ney \& Hatfield (168) called this the ``pseudophotospheric
expansion,'' for the energy distribution and spectroscopic temperature (8,
9, 13) are characteristic of the photosphere of a star with spectral type F
to A. I use here the term ``fireball,'' which has been used to describe the
early development of man-made atomic explosions (see 185, especially
photograph no. 127), to describe this expanding pseudo-photosphere.
Because the fireball is optically thick during its early expansion, it is a
partial calorimeter of the photon luminosity of the embedded remnant,
and the angular expansion rate of the fireball can be combined with
Doppler expansion velocities to obtain the distance to the nova.'' 
He also added ``The angular size of an optically thick fireball
expanding at constant velocity will increase linearly with time regardless
of fluctuations in the luminosity of the central engine, and the angular
expansion can be extrapolated backward to determine its time of origin
(63, 82, 95).''

\begin{figure*}
\epsscale{1.0}
\plottwo{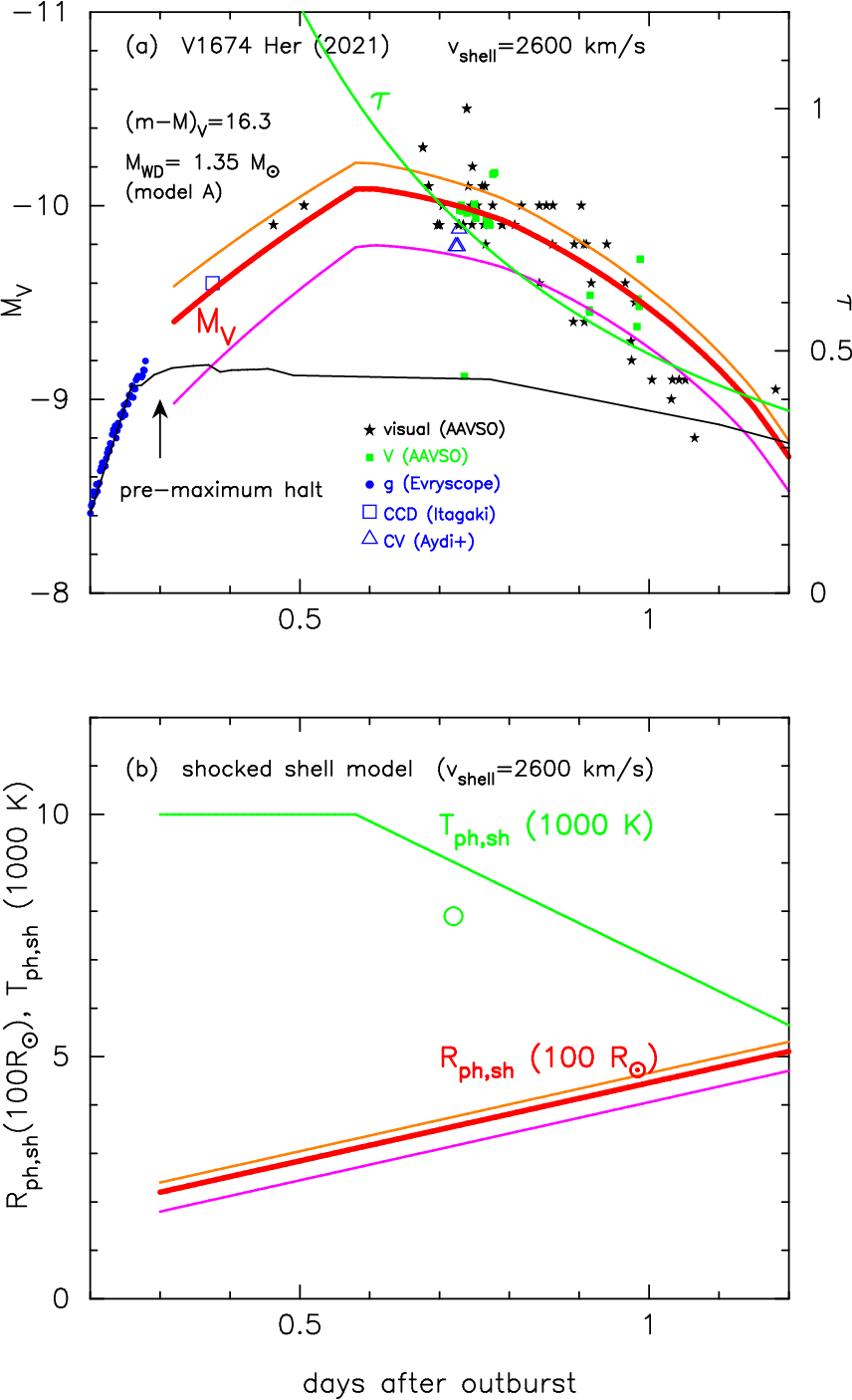}{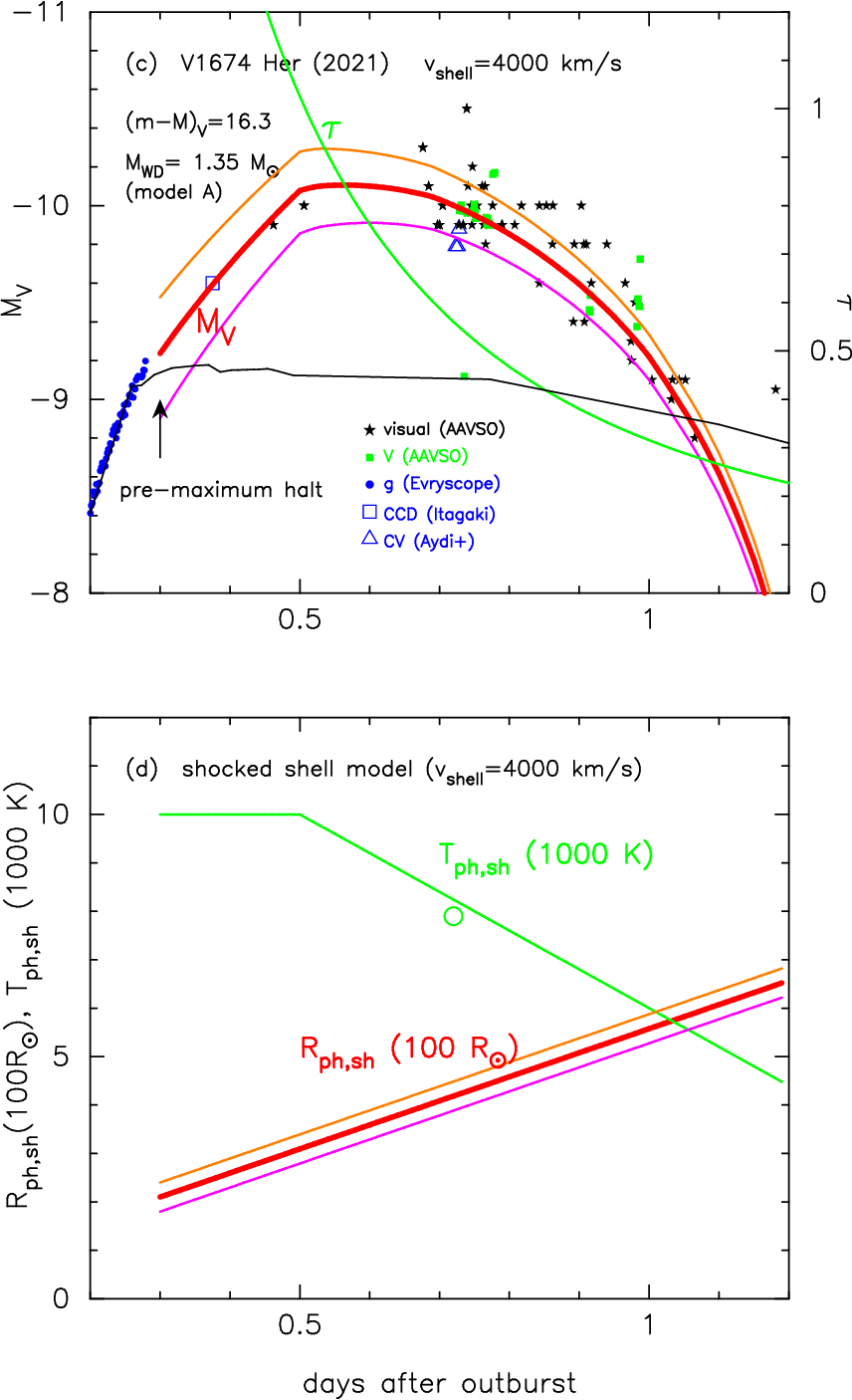}
\caption{(a) The evolution of absolute $V$ brightness, $M_V$, of the shocked
shell near optical maximum in V1674 Her.  The optical data are the same
as those in Figure \ref{v1674_her_v_x_observation_only_logscale}, but
we added the CV magnitudes (open blue triangles) taken from \citet{ayd21sc}.
The orange, thick red, and magenta lines represent
different initial radii of the shocked shell models having the same 
temperature in panel (b), respectively.
We also add the optical depth $\tau$ (green line) of the shocked shell
for the shell mass of $M_{\rm shell}= 3\times 10^{-6} ~M_\sun$.
(b) The assumed evolutions of the photospheric temperature (green line) 
and radii (orange, thick red, and magenta lines)
for our shocked shell models of V1674 Her, mimicking the evolutions
of V339 Del \citep{sko14dt} as in Figure 
\ref{v1674_her_v339_del_v_skopal_wd_photo}(b). 
The green circle denotes the observed
color temperature of $(B-V)_0=0.169$ on day 0.72 \citep{mun21vd}.
Here, we assume the same temperature (green line) evolution for
three different initial radii models, each of which starts
at $R_{\rm ph,sh}= 240 ~R_\sun$ (orange line), $220 ~R_\sun$ (thick red line),
and $180 ~R_\sun$ (magenta line) and expands 
with the same velocity of $v_{\rm shell}= 2600$ km s$^{-1}$. 
The best fit light curve (thick red line) for $v_{\rm shell}=$2600
km s$^{-1}$ is represented numerically by $T_{\rm ph, sh} = 10000$ K 
for $t< 0.6$ days whereas $T_{\rm ph, sh} = 10000 - 8000(t-0.6)$ K
for $t\ge 0.6$ days and $R_{\rm ph, sh}= 220 ~R_\sun + 2600{\rm ~km~s}^{-1}
\times (t-0.3)$ days for $t\ge 0.3$ days. 
(c) Same as in panel (a), but for the expansion velocity of
$v_{\rm shell}= 4000$ km s$^{-1}$.
The three $M_V$ lines (orange, thick red, and magenta lines)
correspond to the initial radii of the same color in panel (d).
(d) Same as in panel (b), but for the expansion velocity of
$v_{\rm shell}= 4000$ km s$^{-1}$ with the initial radii of
$R_{\rm ph,sh}= 240 ~R_\sun$ (orange), 
$210 ~R_\sun$ (thick red), and $180 ~R_\sun$ (magenta).
Here, the best fit light curve (thick red line) for $v_{\rm shell}=$4000
km s$^{-1}$ is represented numerically by $T_{\rm ph, sh} = 10000$ K 
for $t< 0.5$ days whereas $T_{\rm ph, sh} = 10000 - 8000(t-0.5)$ K
for $t\ge 0.5$ days and $R_{\rm ph, sh}= 210 ~R_\sun + 4000{\rm ~km~s}^{-1}
\times (t-0.3)$ days for $t\ge 0.3$ days. 
}
\label{v1674_her_optical_peak_linear}
\end{figure*}

We expect the two gamma-ray novae show a similar temperature 
evolution around the peak.
To see the similarity, we compare 
the effective temperature of V1674 Her on day 0.72, 
$T_{\rm eff}=7800$ K. This corresponds, in Figure 
\ref{v1674_her_v339_del_v_skopal_wd_photo}(a),
to the phase on UT 2014 August 18.5 of V339 Del of which 
the temperature is $T_{\rm eff}=7500$ K. They are 
roughly consistent with each other, indicating that we can use 
the temperature evolution of V339 Del for V1674 Her by squeeze the timescale
by 7.55.

The broad gray lines indicate our simplified smooth trends
for the temporal developments of each value in Figure
\ref{v1674_her_v339_del_v_skopal_wd_photo}(b), which we use a guideline
for our model $T_{\rm ph,sh}$ and $R_{\rm ph, sh}$.
Note that V339 Del shows a flare-like or spike structure in the $V$ light
curve, $T_{\rm eff}$, and $L_{\rm WD}$ between UT 2014 August 16 and 17.
We regard this flare to be a short timescale phenomenon, and exclude
this part from the global evolution of the photosphere.

From the gray lines of $T_{\rm eff}$ and $R_{\rm WD}$, we deduce that \\
(1) $T_{\rm eff}$ starts from $\sim 10,000$ K and gradually declines
to $\sim 6000$ K and \\
(2) $R_{\rm WD}$ linearly increases from $\sim 60 ~R_\sun$ to
$\sim 300 ~R_\sun$.  \\
These two trends of (1) and (2) are also seen in another superbright
Galactic nova V1500 Cyg, as will be shown in Section
\ref{optically_thick_shocked_shell}.
Note that the photospheric radius of V1674 Her could be much larger
than that of V339 Del because its luminosity ($M_{V,\rm max}= -10.2$)
is much brighter than
that of V339 Del \citep[$M_{V,\rm max}= -7.8$, ][]{hac24km}.

Based on these trends, we assume the temperature evolution
as shown in Figure \ref{v1674_her_optical_peak_linear}(b) and (d).
We also assume the expansion velocity of the shock photosphere,
$v_{\rm shell}=$2600 km s$^{-1}$ and 4000 km s$^{-1}$,
with three initial radii of the photosphere when the shock arises.
Thus, the radius of the shocked shell photosphere increases with time
as shown in each panel.

Using these radius $R_{\rm ph,sh}$ and temperature
$T_{\rm ph,sh}$, we have calculated the blackbody luminosity of the
photosphere of the shocked shell, and obtained absolute $V$ magnitude
assuming the bolometric correction for supergiants \citep[e.g.,][]{bohm92}.

The resultant light curves are shown in  
Figure \ref{v1674_her_optical_peak_linear}(a) and (c). 
The thick red line in each panel of (a) and (c) shows a best fit light curve
among the three cases in each panel of (b) and (d).
Their $V$ peaks (of the thick red line) are delayed by $\sim$0.3 days
from the $V$ peak of the free-free emission model light curve (black line).

Our simplified models reproduce the $V$ or visual light curve of
V1674 Her near the optical peak with the observed velocities of
2600--4000 km s$^{-1}$.  This consistency supports our simplified
shocked shell model.  

For comparison, we plot the observed color temperature 
of $(B-V)_0=0.169$ (7800 K) in Figure \ref{v1674_her_optical_peak_linear}(b)
and (d) by a green circle. 
This color temperature is slightly below our green line of assumed
temperature.  Later in Section \ref{optically_thick_shocked_shell} for
V1500 Cyg, we show that the color temperatures are almost consistent
with our assumed green line.
If we decrease the green line down
from 8500 K (original position) to 7800 K (green circle)
in Figure \ref{v1674_her_optical_peak_linear}(b),
then the photospheric radius should be increased by a factor of
$(8500/7800)^2=1.19$ to keep the luminosity to be the same.
In Figure \ref{v1674_her_optical_peak_linear}(d),
the photospheric radius should be increased by a factor of
$(8000/7800)^2=1.05$, where we use the blackbody luminosity of
$L_{\rm ph}= 4\pi R_{\rm ph}^2 \sigma  T_{\rm ph}^4$ and 
$\sigma$ is the Stefan-Boltzmann constant.

\subsection{Optical depth of the shocked shell}
\label{optical_depth_shock}

Here, we estimate the optical depth of the shocked shell.
The optical depth $\tau_{\rm shell}$ is approximately calculated from
\begin{equation}
\tau_{\rm shell} \equiv \int_{\rm shell} \kappa \rho d r 
 \approx  {\kappa {M_{\rm shell}} \over {4\pi R^2_{\rm shell}}},
\label{potical_depth_shell}
\end{equation}
where $\kappa$ is the opacity, $\rho$ the density, $r$ the radius from
the center of the WD, and $M_{\rm shell}$ the mass of, and 
$R_{\rm shell}$ the radius of the shocked shell. 
  
We take the opacity of $\kappa\sim 1$ g$^{-1}$ cm$^2$.
The radius of the shocked shell is assumed to be the same as the
photospheric radius shown in Figure \ref{wind_shock_config},
i.e., $R_{\rm shell}\approx R_{\rm ph,sh}$.  
More exactly, we use the photospheric radius of the red line
that starts from $R_{\rm ph,sh}=220 ~R_\sun$ with $v_{\rm shell}=2600$ km
s$^{-1}$ in Figure \ref{v1674_her_optical_peak_linear}(a), or from
$R_{\rm ph,sh}=210 ~R_\sun$ with $v_{\rm shell}=4000$ km s$^{-1}$
in Figure \ref{v1674_her_optical_peak_linear}(c).

The most important unknown parameter is the mass of the shocked shell.
The ejecta mass was observationally obtained to be from
$M_{\rm ej}\sim$(3--7)$\times 10^{-5} ~M_\sun$ \citep{hab24},
$2\times 10^{-5}$--$2\times 10^{-4}~M_\sun$ \citep{dra21}, 
to $1.4$ (0.2--2.2) $\times 10^{-3} ~M_\sun$ \citep{woo21}.
If we adopt $M_{\rm shell}\sim 1\times 10^{-3} ~M_\sun$ as an upper limit,
our model light curves of the shocked shell cannot reproduce 
the $V$ light curve of V1674 Her around the optical $V$ peak.
This is simply because the optical depth of the shocked shell 
remains optically thick for a long time as shown below. 


We estimate the day ($t$) when the optical depth becomes smaller than
unity ($\tau < 1$) with Equation (\ref{potical_depth_shell}) 
assuming the expansion velocity of the shell $v_{\rm shell}= 4000$ km
s$^{-1}$ and $R_{\rm shell}= v_{\rm shell}\times t$.
If we adopt $M_{\rm shell}\sim 1\times 10^{-3} ~M_\sun$, we found that
the optical depth becomes $\tau < 1$ on day 11.5 at
$R_{\rm shell}=5700~R_\sun$ for the expansion
velocity of 4000 km s$^{-1}$.  The $V$ brightness becomes much fainter
than that of the free-free emission model light curve after it crosses
the black line in Figure \ref{v1674_her_optical_peak_linear}.
On the other hand, if we use $M_{\rm shell}\sim 3\times 10^{-6} ~M_\sun$
as shown below, we obtain day 0.63 ($\tau < 1$)
at $R_{\rm shell}=314~R_\sun$
for the same expansion velocity of 4000 km s$^{-1}$.
The day 0.63 is close to the epoch of optical $V$ maximum, after which
the $V$ brightness starts to rapidly decreases as shown in Figure
\ref{v1674_her_optical_peak_linear}.  In this case, the shocked shell
becomes optically thin and the free-free emission brightness replaces
that of the shocked shell.  Thus, the shell mass should be as small as
$M_{\rm shell}\sim 3\times 10^{-6} ~M_\sun$ in order to reproduce
the $V$ light curve around the optical peak of V1674 Her.

In theoretical model calculations, \citet{kat25hs} listed the ignition
mass of their model A ($1.35 ~M_\sun$ WD with the mass accretion rate
of $\dot{M}_{\rm acc}=1\times 10^{-11} ~M_\sun$ yr$^{-1}$) to be
$1.6 \times 10^{-6} ~M_\sun$.  However, Kato et al.'s explosion model
did not include the core material mixing process.  This corresponds to
the lowest limit of the ejecta mass.
\citet{yar05} listed the ejecta mass
to be $\lesssim 3\times 10^{-6}~M_\sun$ for $1.4 ~M_\sun$ WDs with 
$\dot{M}_{\rm acc}= 1\times 10^{-11} ~M_\sun$ yr$^{-1}$,
which is 1.3--2.2 times the accreted mass.
Therefore, we double our ignition mass of $1.6\times 10^{-6} ~M_\sun$
and adopt $M_{\rm shell}= 3 \times 10^{-6} ~M_\sun$ to estimate
the optical depth of the shell in 
Figure \ref{v1674_her_optical_peak_linear}(a) and (c).

The green line in Figure \ref{v1674_her_optical_peak_linear}(a) presents
$\tau_{\rm shell}$ for the expansion velocity of $v_{\rm shell}=2600$ km
s$^{-1}$.  The optical depth $\tau_{\rm shell}$ continuously decreases
and becomes below $\tau_{\rm shell}=1$ on day 0.6, where we start the
decrease in the photospheric temperature from $T_{\rm ph,sh}=10,000$ K
to 5000 K after the decay trend in Figure 
\ref{v1674_her_v339_del_v_skopal_wd_photo}(b).    
For the case of $v_{\rm shell}=4000$ km s$^{-1}$ in Figure
\ref{v1674_her_optical_peak_linear}(d), 
we assume the same decay trend of the photospheric temperature as in
Figure \ref{v1674_her_optical_peak_linear}(b).
The optical depth $\tau_{\rm shell}$ continuously decreases and becomes
below $\tau_{\rm shell}=1$ on day 0.5.

\begin{figure*}
\epsscale{1.0}
\plotone{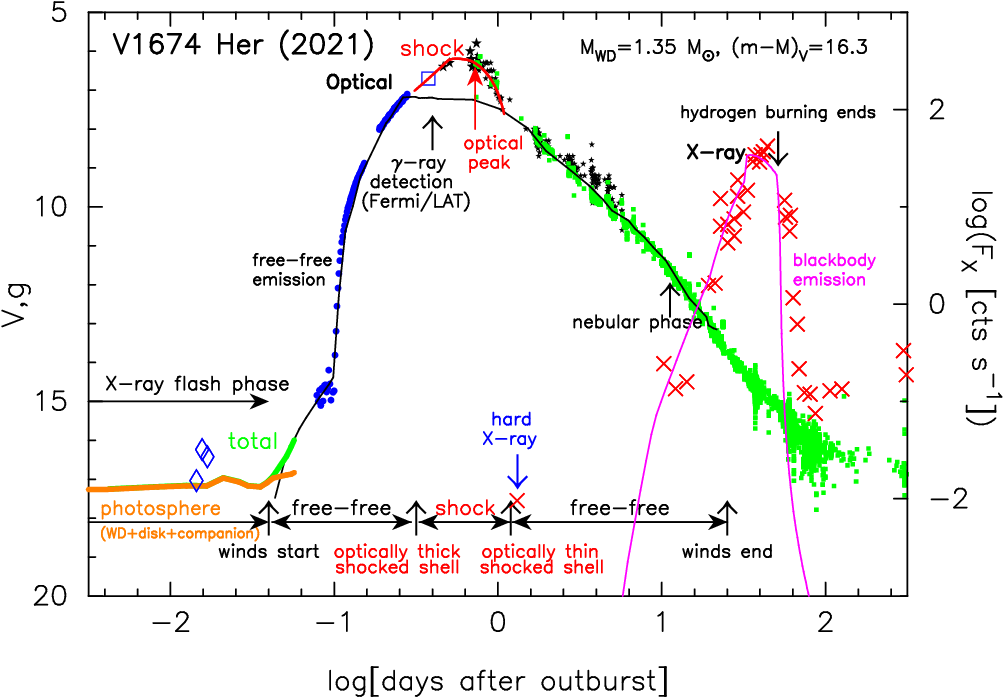}
\caption{
Same as Figure \ref{v1674_her_v_x_observation_only_logscale}, but
we added $V$ light curve (red line labeled ``shock'')
for the optically thick shocked shell in Figures
\ref{wind_shock_config}(d) and \ref{v1674_her_optical_peak_linear}(c). 
Thus, the chronological order is a shock formation ($t=0.32$ day,
$1.35 ~M_\sun$ WD model), emergence of gamma-rays ($t=0.39$ day, observation),
and optical $V$ peak ($t=$0.7--0.8 day, observation).
The optically thick shocked shell becomes optically thin on day 1.2, and
we are able to see the high velocity inner wind
and the higher temperature WD photosphere
as illustrated in Figure \ref{wind_shock_config}(c).
The photons from the central WD are all absorbed by the optically thick
shocked shell, so we are not able to observe the nova WD (black line) between
day 0.3 and 1.2. Instead, we observe the optically thick shocked shell
photosphere (red line),
as depicted in Figure \ref{wind_shock_config}(d). 
\label{v1674_her_v_x_observation_only_logscale_no2_final}}
\end{figure*}

The best-fit $M_V$ magnitudes 
for both the $v_{\rm shell}=2600$ km s$^{-1}$ and 4000 km s$^{-1}$ cases
are very similar to each other (thick red lines in
Figure \ref{v1674_her_optical_peak_linear}(a) and (c)). 
We plot the $v_{\rm shell}=$4000 km s$^{-1}$ case in 
Figure \ref{v1674_her_v_x_observation_only_logscale_no2_final} until 
the optical depth of the shell decreases to $\tau_{\rm shell}=0.35$.
This line reasonably fits with the $V$ light curve around the peak. 

We suppose that the large difference between the observed ejecta mass
and theoretical ejecta mass comes from the filling factor of the ejecta gas.
If all the ejecta mass is confined into the shocked shell
\citep[e.g., ][]{hac22k}, its filling factor becomes a tenth (0.1)
or hundredth (0.01) and, as a result, the estimated observed mass is
broadly consistent with the theoretical ejecta masses.
\citet{hab24} adopted the filling factor of 0.1 and gave a tenth of
the other estimates by \citet{woo21} and \citet{dra21}.


\subsection{Transition from optically thick to thin}
\label{from_thick_to_thin}

The optical depth of the shocked shell 
is calculated using Equation (\ref{potical_depth_shell}) in 
Section \ref{optical_depth_shock}. 
It decreases with time and gradually
becomes $\tau < 1$ as shown in Figure 
\ref{v1674_her_optical_peak_linear}(a) and (c).
Here, $\tau$ (or $\tau_{\rm shell}$) is the optical depth of the shocked shell.
We regard that the shocked shell becomes sufficiently optically thin on day 
1.1--1.2, where the $V$ luminosity of the shocked shell photosphere
becomes fainter than that of free-free emission as shown in
Figure \ref{v1674_her_optical_peak_linear}(a) and (c). 

Thus, we conclude that the shocking power dominates the $V$ luminosity
near optical peak between day 0.32 and 1.2.  We see the recombination
front of the shocked shell so that the optical spectra show a simple
P-Cygni profile with the absorption velocity of $\sim 3000$ km s$^{-1}$
during day 0.32--1.2 \citep[see Figure 2(a) of ][for a spectrum
on day 0.747]{hab24}.  

There are slight differences in the velocity among
the observations.  \citet{mun21vd} listed $-3100$ km s$^{-1}$
while \citet{hab24} reported $-3600$ km s$^{-1}$ for H$\alpha$ P-Cygni
profiles.  \citet{mun21vd} also listed $-2700$ km s$^{-1}$ for
P-Cygni profiles of \ion{He}{1} lines.  So, we adopt 2600 km s$^{-1}$
as a lower limit value and 4000 km s$^{-1}$ as an upper limit value.

The velocity of the shocked shell increases but this rate is very 
small as clearly shown by the shock calculation in Figure 1 of \citet{hac22k}.
Therefore, we assume a constant shocked shell velocity in our modeling
for the optically thick shocked shell.  This assumption is also
supported by the time series of the P-Cygni profiles mentioned above.

\subsection{Parameter dependence of light curves of the recombination front}
\label{light_curve_recombination}

Based on the above two trends (1) and (2),
we adopt the three cases of the shocked shell photospheric radius
$R_{\rm ph,sh}$ evolution and one case of the shocked shell photospheric
temperature $T_{\rm ph,sh}$ evolution in Figure 
\ref{v1674_her_optical_peak_linear}(b)
for the expansion velocity of $v_{\rm shell}= 2600$ km s$^{-1}$,  
and calculate $M_V$ light curves of V1674 Her
in Figure \ref{v1674_her_optical_peak_linear}(a).
Here, $R_{\rm ph,sh}$ begins to start at $R_{\rm ph,sh}= 240 ~R_\sun$
(orange line), $220 ~R_\sun$ (thick red line), and $180 ~R_\sun$
(magenta line) on day 0.3 and expands with the velocity of 
$v_{\rm shell}= 2600$ km s$^{-1}$.
Among the three models, the thick red line (starting from $220 ~R_\sun$)
is best fit with the observation in Figure
\ref{v1674_her_optical_peak_linear}(a).

Changing the expansion velocity to $v_{\rm shell}= 4000$ km s$^{-1}$,
we obtain similar $M_V$ light curves as shown
in Figure \ref{v1674_her_optical_peak_linear}(c).  
Here, the photospheric radius begins to start at $R_{\rm ph,sh}= 240 ~R_\sun$
(orange), $210 ~R_\sun$ (thick red), and $180 ~R_\sun$ (magenta) on day 0.3
and expands with the velocity of $v_{\rm shell}= 4000$ km s$^{-1}$ as shown
in Figure \ref{v1674_her_optical_peak_linear}(d).
Among these three, the thick red line (starting from $210 ~R_\sun$)
is best fit with the observation in Figure
\ref{v1674_her_optical_peak_linear}(c).
We plot this best fit model (red line) also in Figure 
\ref{v1674_her_v_x_observation_only_logscale_no2_final}.
For the two expansion velocities of 2600 and 4000 km s$^{-1}$,
both the best fit model light curves (thick red lines) evolve similarly.
Thus, we conclude that we are able to reproduce the $V$ light curve of
V1674 Her around the optical peak by our optically thick shocked shell model.

\subsection{Chronological order of shock, gamma-ray, and optical maximum}
\label{chronological_order}

Theoretically, our model $V$ light curves (thick red lines)
reach maximum of $M_V= -10.1$ ($V=6.2$) on day 0.5--0.6 
(Figure \ref{v1674_her_optical_peak_linear}(a) and (c)).
The WD photospheric emission and free-free emission are
obscured by the optically thick shocked shell and not directly observed
from the Earth.  Observationally, the $V$ magnitude seems to attain
its maximum on day $\sim$0.7, although we do not know the exact $V$ maximum
of V1674 Her because there are no visual or $V$ data
between day 0.5 and 0.7.
The absolute $V$ magnitude of $M_V= -10.1$ requires $R_{\rm ph,sh}\sim 300
~R_\sun$ for $T_{\rm ph,sh}=10000$ K.  This is roughly consistent with
the expansion velocity of $\sim$3000 (4000) km s$^{-1}$ and
$t_{V,\rm max}\sim$ 0.8 (0.6) days because 
$R_{\rm shell}\approx v_{\rm shell}\times t_{V,\rm max} \approx$
3000 (4000) km s$^{-1}\times$ 0.8 (0.6) days$\approx 300 ~R_\sun$,
where $t_{V,\rm max}$ is the epoch of maximum $V$ light. 

Then, the chronological order of events are 
the maximum expansion of the WD photosphere 
and shock-arising on day 0.32, 
emergence of GeV gamma rays on day 0.39 \citep{sok23},
and optical maximum on day $\sim$0.7 (observationally), as plotted
in Figures \ref{v1674_her_optical_peak_linear}
and \ref{v1674_her_v_x_observation_only_logscale_no2_final}.


\begin{figure*}
\epsscale{0.75}
\plotone{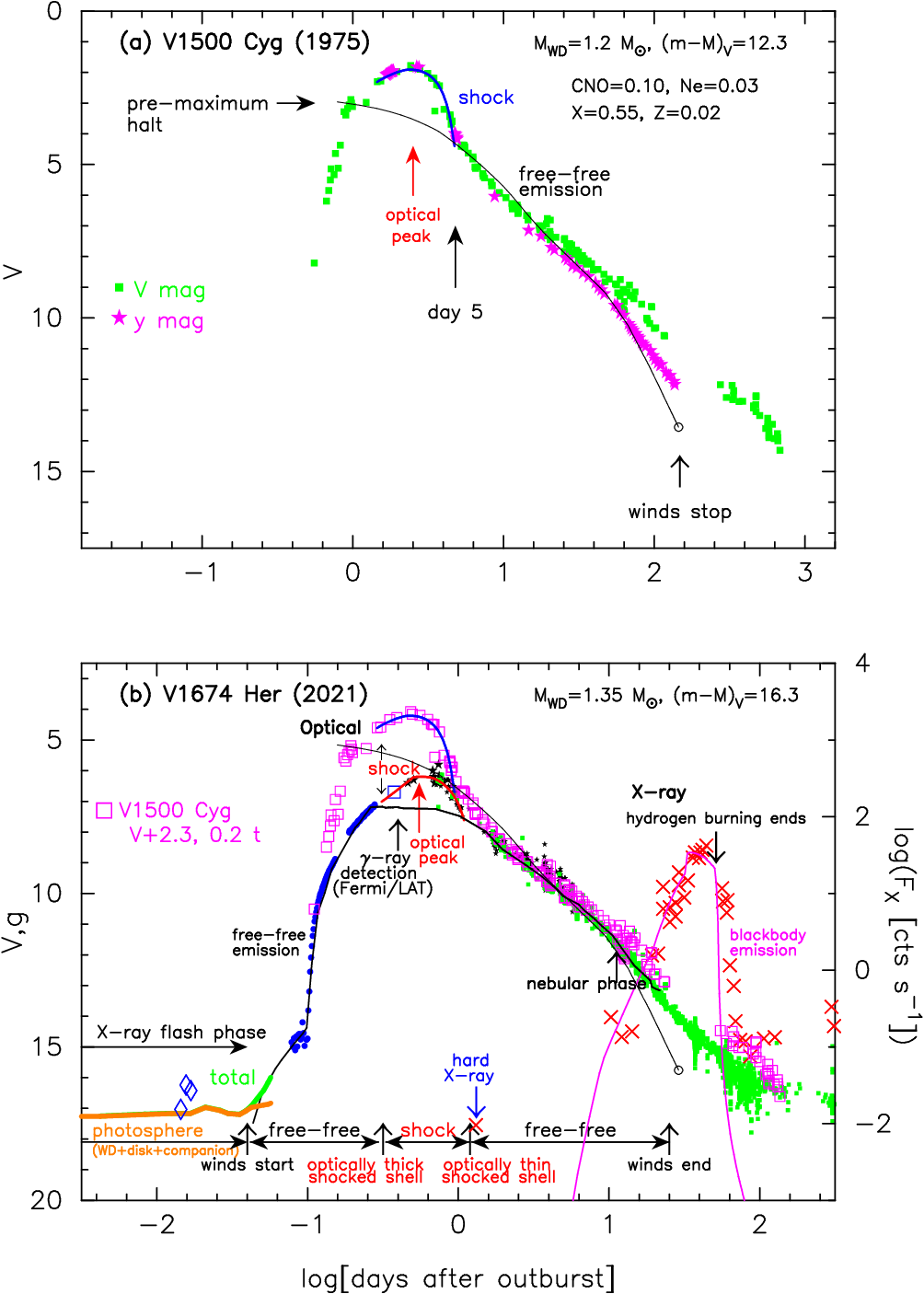}
\caption{
(a) The $V$ and $y$ light curves of V1500 Cyg are plotted against
a logarithmic time, days after outburst.  The outburst day is assumed to
be $t_{\rm OB}=$JD 2442653.0 $=$ UT 1975 August 28.5.  The $V$ data
(filled green squares) are taken from \citet{tem79} while the $y$ magnitudes
(filled magenta stars) are from \citet{lock76m}.
The free-free emission model light curve (thin black line)
of $1.2 ~M_\sun$ WD (Ne2) is taken from \citet{hac10k, hac14k}.
We calculated the shocked shell model light curve (blue line) in 
Section \ref{optically_thick_shocked_shell},
assuming the initial $R_{\rm ph,sh}= 280 ~R_\sun$ on day 1.4 
and the expansion velocity of $v_{\rm shell}= 1700$ km s$^{-1}$
\citep{bol76g, feh76a}.
(b) Same as Figure \ref{v1674_her_v_x_observation_only_logscale_no2_final},
but we overlap the $V$ light curve of V1500 Cyg (open magenta squares)
with that of V1674 Her by 2.3 mag down and 5 times squeeze of time
as denoted by ``V1500 Cyg V+2.3, 0.2 t.'' 
See Section \ref{sec_timestretching} for details.
\label{v1500cyg_1674her_model_v_observation_logscale}}
\end{figure*}

\begin{figure}
\epsscale{1.15}
\plotone{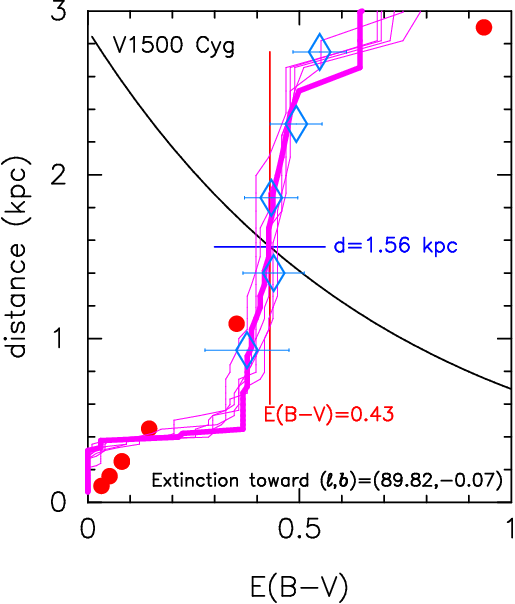}
\caption{
The distance-reddening relations toward V1500 Cyg whose galactic coordinates
are $(\ell, b)= (89\fdg82, -0\fdg07)$.
The black line denotes the relation of Equation (\ref{distance_reddening_law})
together with $(m-M)_V=12.3$ for V1500 Cyg.  The thin magenta lines are
the sample distance-reddening relations given by \citet{gre19}
while the thick magenta line is their best-fit line for them.
Here, we use the relation of $E(B-V)= 0.884\times$(Bayestar19)
\citep[see the Bayestar website of][]{gre19}.
The two relations (black and magenta lines) cross at the distance of
$d=1.56$ kpc and $E(B-V)=0.43$.
The filled red circles denote the distance
and reddening of nearby stars given by \citet{you76}.
The unfilled cyan-blue diamonds with error bars
represent the relation of \citet{ozd18}.
}\label{distance_reddening_v1500_cyg_xxx_no2}
\end{figure}

\begin{figure*}
\epsscale{1.0}
\plotone{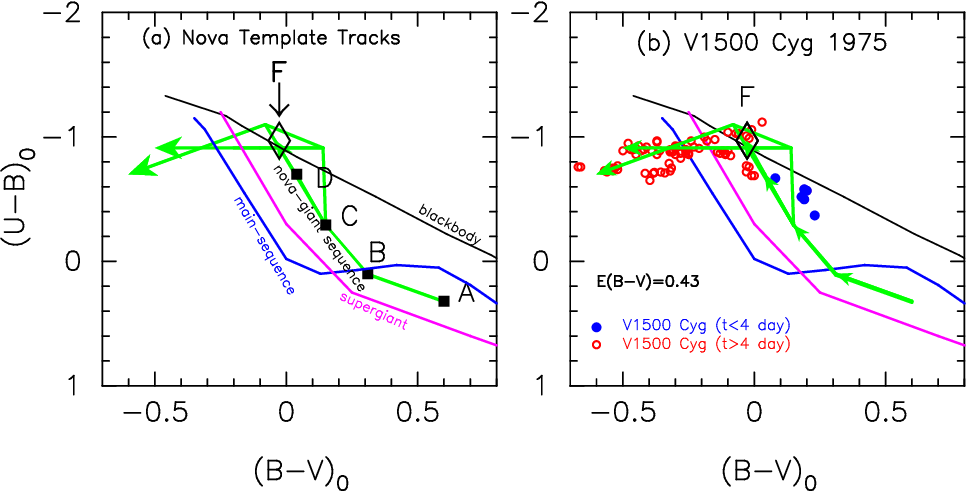}
\caption{
Color-color evolution of classical novae
in the intrinsic $(B-V)_0$-$(U-B)_0$ diagram.
Green lines and arrows: nova template tracks (evolution paths)
taken from \citet{hac14k}. 
(a) Dereddened color-color diagram of FH~Ser on the nova-giant sequence:
four evolutionary stages are specified by A, B, C, and D
beside the filled black squares.
These color data are tabulated in Table 1 of \citet{hac14k}.
Open diamond labeled F indicates the landmark for optically-thick
free-free emission spectra.
%
(b) Color-color evolution of V1500~Cyg 1975. 
Data are same as those in \citet{hac14k} and dereddened with $E(B-V)=0.43$ 
(see Section \ref{sec_timestretching}).
The data are separately denoted by filled blue
($t< 4$ days) and open red ($t>4$ days) circles. 
\label{color_color_diagram_template_v1500_cyg_2fig}}
\end{figure*}

\section{Light curve of V1500 Cyg}
\label{light_curve_v1500_cyg}

\citet{del91} listed V1500 Cyg as a superbright nova in our Galaxy (see
also Figure \ref{vmax_t2_selvelli2019_schaefer2018_2fig}).
Here, we compare V1500 Cyg with V1674 Her and deduce common properties
among the two superbright novae.

Figure \ref{v1500cyg_1674her_model_v_observation_logscale}(a) shows the
$V$ (filled green squares) and $y$ (filled magenta stars) light curves
of V1500 Cyg against a logarithmic time as well as
the shocked shell model light curve (blue line) that will be calculated
in Section \ref{optically_thick_shocked_shell}. 
This figure also shows the free-free emission
model light curve (thin black line) based on the steady-state nova wind
solutions \citep{kat94h, hac06kb, hac14k}, where
we adopt the outburst day of $t=0= t_{\rm OB}=$JD 2442653.0$=$
UT 1975 August 28.5 after \citet{enn77}, and
the $V$ band distance modulus of $(m-M)_V=12.3$ 
after \citet{hac14k}.  This $V$ band distance modulus
is consistent with the extinction of $E(B-V)=0.45$ \citep[e.g.,
][]{tom76wl, you76}
and the Gaia eDR3 distance of   
$d= 1567^{+270}_{-192}$ pc \citep{bai21rf} together with the relation of
\begin{equation} 
(m-M)_V=3.1 E(B-V) + 5 \log (d/{\rm 10 ~pc}),
\label{distance_reddening_law}
\end{equation}
where $E(B-V)$ is the extinction and $d$ is the distance
toward V1500 Cyg, as shown in Figure 
\ref{distance_reddening_v1500_cyg_xxx_no2}.

\subsection{Time-Stretching Method}
\label{sec_timestretching}

In this subsection, we determine the distance modulus to V1674 Her
with the time-stretching method, which is a powerful way
to obtain the distance modulus in the $V$ band, $(m-M)_V$,
toward a nova, and has ever been applied to a number of novae
\citep{hac10k, hac15k, hac16k, hac18k, hac25kv392per, hac20skhs,
hac24km, hac25kw, kat25hs}.

Nova light curves often show a common property; if two nova light curves
are plotted in the logarithmic time and shift in the vertical and horizontal
directions, the major part of these light curves are overlapped
each other independently of the WD mass, chemical composition, and speed
class of novae \citep{hac06kb, hac20skhs}.
Using this remarkable property, we can determine
the distance to a nova (target nova: V1674 Her) by comparing
a well studied nova with known distance (template nova: V1500 Cyg).

Here, we describe the $V$ light curves of the target nova as
$(m[t])_{V,\rm target}$ and the template nova $(m[t])_{V,\rm template}$.
When we adopt an appropriate time-stretching parameter $f_{\rm s}$,
these two nova $V$ light curves overlap each other.
We shift the template nova light curve in the horizontal direction
by a factor of $f_{\rm s}$ in the logarithmic scale
($t \rightarrow t\times f_{\rm s}$),
and move vertically down by $\Delta V$. This vertical shift
can be written as
\begin{equation}
(m[t])_{V,\rm target} = \left((m[t \times f_{\rm s}])_V
+ \Delta V\right)_{\rm template}.
\label{overlap_brigheness}
\end{equation}
As the two nova light curves are overlapping,
their distance moduli in the $V$ band satisfy
\begin{eqnarray}
& & (m-M)_{V,\rm target} \cr
&=& ( (m-M)_V + \Delta V )_{\rm template} - 2.5 \log f_{\rm s}.
\label{distance_modulus_formula}
\end{eqnarray}
Here, $m_V$ and $M_V$ are the apparent and absolute $V$ magnitudes,
and $(m-M)_{V, \rm target}$ and $(m-M)_{V, \rm template}$ are
the distance moduli in the $V$ band
of the target and template novae, respectively.
\citet{hac18k, hac18kb, hac19k, hac19kb, hac21k} confirmed that
Equations (\ref{overlap_brigheness}) and (\ref{distance_modulus_formula})
are also broadly valid for other $U$, $B$, and $I$ (or $I_{\rm C}$) bands.

This remarkable similarity is demonstrated in
Figure \ref{v1500cyg_1674her_model_v_observation_logscale}(b), which
compares the $V$ light curve of V1674 Her with V1500 Cyg.
These two novae are well overlapped to each other if we squeeze
the timescale of V1500 Cyg by 5 times and shift down the $V$ magnitude
by 2.3 mag as labeled ``V1500 Cyg V+2.3, 0.2 t.''
It should be noted that we try to overlap the post-maximum phase that 
follows the universal decline law ($L_V\propto t^{-1.75}$ line in Figure
\ref{v1674_her_v_x_observation_only_logscale})
as long/much as possible. 

In Figure \ref{v1500cyg_1674her_model_v_observation_logscale}(b),
we regard V1674 Her as the target and V1500 Cyg as the template
in Equation (\ref{overlap_brigheness}).
As V1500 Cyg evolves 5 times slower, we adopt $f_{\rm s}= 0.2$
and $\Delta V= +2.3$ and have the relation of
\begin{eqnarray}
(m&-&M)_{V, \rm V1674~Her} \cr
& = & (m - M + \Delta V)_{V, \rm V1500~Cyg} - 2.5 \log 0.2 \cr
&=& 12.3 + 2.3\pm 0.2 + 1.75 = 16.35\pm 0.2,
\label{distance_modulus_v1674_her_lv_vul_v}
\end{eqnarray}
where we adopt $(m-M)_{V, \rm V1500~Cyg}=12.3$ after \citet{hac14k}.
This result of $(m-M)_{V, \rm V1674~Her}=16.35\pm 0.2$
is the same as that of \citet{kat25hs}, obtained
with the same time-stretching method but against other three template novae,
LV Vul, V339 Del, and KT Eri.

Figure \ref{distance_reddening_v1500_cyg_xxx_no2} shows
the distance-reddening relation (black line) 
calculated by Equation (\ref{distance_reddening_law})
together with $(m-M)_{V, \rm V1500~Cyg}=12.3$.
This black line crosses \citet{gre19}'s relation (magenta lines)
at the distance of $d=1.56$ kpc and the reddening of 
$E(B-V)=0.43$.  
Here, we use the relation of $E(B-V)= 0.884\times$(Bayestar19)
\footnote{http://argonaut.skymaps.info} given by \citet{gre19}. 
These two values are consistent each with
the Gaia eDR3 distance of
$d= 1567^{+270}_{-192}$ pc \citep{bai21rf}
and the extinction of $E(B-V)=0.45$ obtained by \citet{tom76wl}.


\subsection{Another Galactic superbright nova V1500 Cyg}
\label{comparison_v1500_cyg}

\citet{gal76} obtained the V1500 Cyg brightnesses for the three broad optical
$V$, $R$, and $I$ bands and the eight infrared 1.2, 1.6, 2.2, 3.6, 4.8,
8.5, 10.6, and $12.5~\mu$m bands during the 50 days following the discovery.
They estimated the outburst day to be 
UT 1975 August 28.9$(=$JD 2442653.4) from the data
of angular expansion of the pseudo-photosphere.
They concluded that the spectral energy distribution is approximately
that of a blackbody (blue line in Figure
\ref{v1500cyg_1674her_model_v_observation_logscale}(a))
during the first 3 days while it is close to
$F_\nu =$~constant after the fourth day, where $F_\nu$ is the flux
at the frequency $\nu$.  This $F_\nu =$~constant
spectra resemble those usually ascribed to the free-free emission
(black line in Figure \ref{v1500cyg_1674her_model_v_observation_logscale}(a)).

\citet{enn77} obtained similar results, but based on the infrared
photometry from 1 to $20~\mu$m.  The nova spectrum changed from
a blackbody to a bremsstrahlung emission at day $\sim 4-5$,
that is, from that of a Rayleigh-Jeans tail ($F_\nu \propto \nu^2$)
to that of a thermal bremsstrahlung emission ($F_\nu \sim$ constant).

They also obtained the onset of outburst on $t_{\rm OB}=$JD 2442653.0$\pm0.5$
from an analysis of the photospheric expansion similar to that
by \citet{gal76}.  Therefore, we define the outburst day
of V1500 Cyg as $t=0=t_{\rm OB}=$JD 2442653.0 in our plot in
Figure \ref{v1500cyg_1674her_model_v_observation_logscale}(a).

The transition of the shocked shell, from optically thick to 
thin, can be confirmed from the evolution in the color-color diagram. 
Figure \ref{color_color_diagram_template_v1500_cyg_2fig}(a) shows
a dereddened color-color $(B-V)_0$-$(U-B)_0$ diagram. 
A typical nova evolves to follow the nova template tracks 
(thick green arrows) as demonstrated by \citet{hac14k}.  
The evolution of V1500 Cyg is shown in 
Figure \ref{color_color_diagram_template_v1500_cyg_2fig}(b) 
assuming the color excess of $E(B-V)=0.43$ based on the
result in Section \ref{sec_timestretching}. 
In the rising and near peak phase ($t<4$ days, blue dots),
the positions of V1500 Cyg are close to the blackbody sequence
(black line) apart from the nova giant sequence (green line).
After the optical peak ($t>4$ days, open red circles), it approaches
point F (i.e., free-free emission) and then moves almost horizontally 
leftward along the typical nova template tracks (green arrows).

To summarize, the nova spectrum is close to that of the blackbody around
the optical peak, and then, about 5 days after the outburst, it
enters a phase in which free-free emission dominates.
In other words, these observations can be interpreted as 
the detection of the transition from optically thick photosphere
of the shocked shell (blue line in
Figure \ref{v1500cyg_1674her_model_v_observation_logscale}(a))
to free-free emission (thin black line in
Figure \ref{v1500cyg_1674her_model_v_observation_logscale}(a))
coming from much inner region close to the WD photosphere.
This transition is essentially the same as
that of V1674 Her on day $\sim 1$, as shown in Figure 
\ref{v1674_her_v_x_observation_only_logscale_no2_final}.
These transitions accompany sharp drops from the $V$ peak both for V1674 Her
and V1500 Cyg.   This kind of luminosity drops are also observed in
the recombination front of hydrogen as seen in shock-heated expanding
SNe IIP ejecta \citep[e.g.,][]{dub25}.

Thus, we regard that the two superbright novae, V1500 Cyg and V1674 Her,
have an optically thick shocked shell around the optical peak, 
which makes them superbright novae.


\begin{figure}
\epsscale{1.0}
\plotone{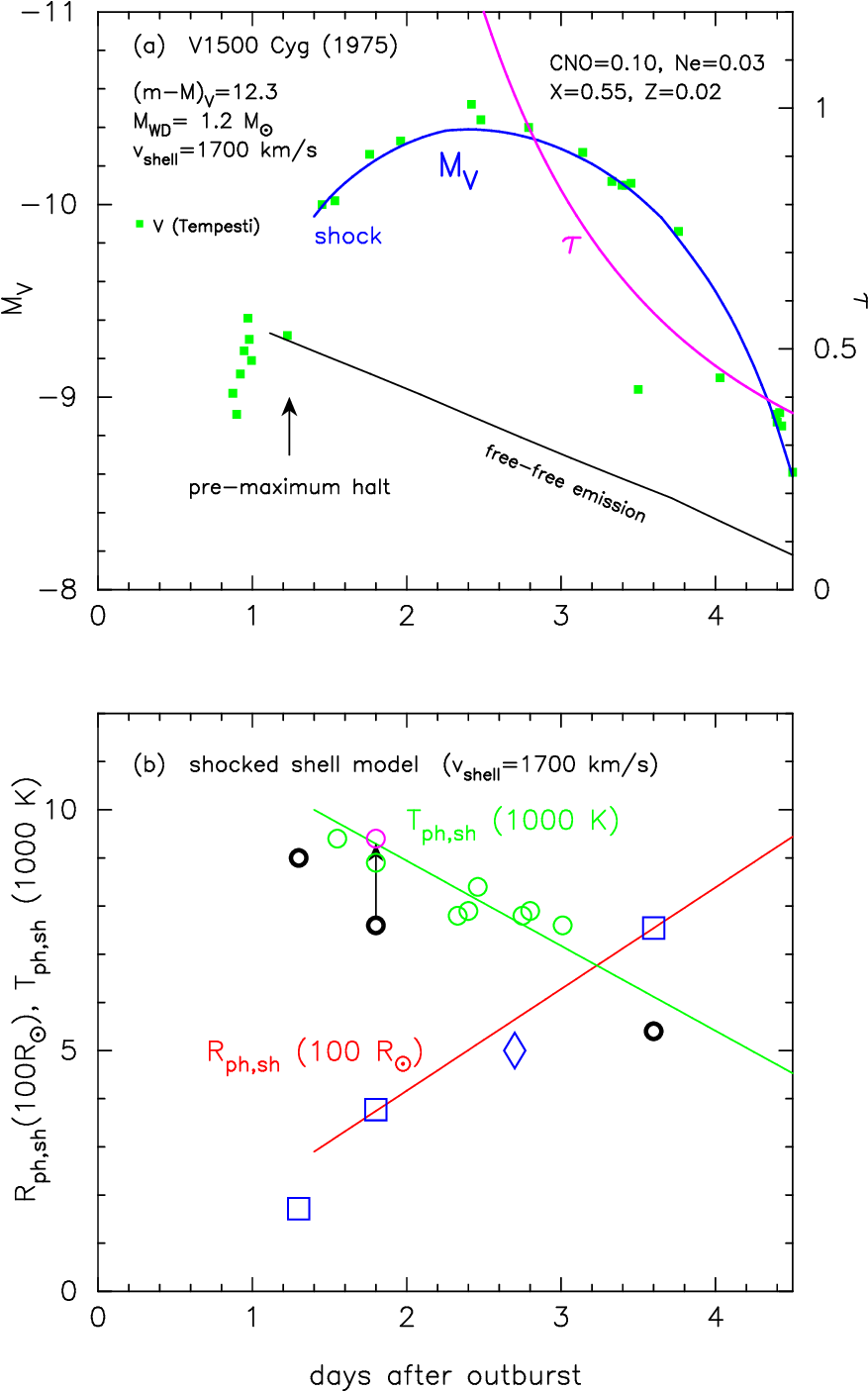}
\caption{
(a) Same as Figure \ref{v1674_her_optical_peak_linear}(a), but for 
V1500 Cyg.  The observed $V$ magnitudes are the same as those in
Figure \ref{v1500cyg_1674her_model_v_observation_logscale}(a).
The evolution of absolute $V$ brightness, $M_V$, 
of the shocked shell (blue line) is calculated from
the photospheric temperature $T_{\rm ph,sh}$ and radius $R_{\rm ph,sh}$ 
in panel (b) assuming the blackbody luminosity with the bolometric
correction of supergiants.  
We added the free-free emission model $V$ light curve (black line).
We also add the optical depth $\tau$ (magenta line) 
for the shocked shell for the 
shell mass of $M_{\rm shell}= 3\times 10^{-6} ~M_\sun$.
(b) The evolutions of photospheric temperature $T_{\rm ph,sh}/1000$ K
(green line) and radii $R_{\rm ph,sh}/100~R_\sun$ (red line)
for our shocked shell models.
Here, we assume that the radius starts on day 1.4 from
$R_{\rm ph,sh}= 290 ~R_\sun$ and expands 
with the velocity of $v_{\rm shell}= 1700$ km s$^{-1}$. 
The best fit light curve (blue line in panel (a))
for $v_{\rm shell}=$1700 km s$^{-1}$ is represented numerically
by $T_{\rm ph, sh} = 10000 - (5500/3.2)(t-1.4)$ K
for $t\ge 1.4$ days and 
$R_{\rm ph, sh}= 290 ~R_\sun + 1700{\rm ~km~s}^{-1}
\times (t-1.4)$ days for $t\ge 1.4$ days in panel (b). 
The open blue diamond is an estimate of $R_{\rm ph,sh}\sim 500 ~R_\sun$
on day 2.7 near optical $V$ maximum by \citet{fer86lw}. 
We also added the temperatures (open black circles) and radii (open blue
squares) estimated by \citet{gal76}.
The open magenta circle pointed by the upward arrow indicates
the temperature corrected with the true brightness of $M_V\approx -10.2$.
The blackbody temperatures (open green circles) are
calculated with the intrinsic $(B-V)_0$ color reported in IAUC
2826, 2828, and 2830.
}
\label{v1500_cyg_optical_peak_linear}
\end{figure}

\subsection{Optically thick shocked shell in V1500 Cyg}
\label{optically_thick_shocked_shell}

Figure \ref{v1500cyg_1674her_model_v_observation_logscale}(b) 
shows a remarkable agreement in the $V$ light curves of V1674 Her
and V1500 Cyg except for the optical maximum phase. 
In the post-maximum phase, free-free emission dominates the optical flux
and the light curve follows the universal decline law 
($L_V\propto t^{-1.75}$). 
This kind of resemblance among the two nova light curves 
has been theoretically explained by \citet{hac06kb}. 

In this subsection, we calculate the light curve around the optical peak
of V1500 Cyg which seems to be much brighter than V1674 Her
in Figure \ref{v1500cyg_1674her_model_v_observation_logscale}(b). 
We use our optically-thick shocked-shell model. 

Figure \ref{v1500cyg_1674her_model_v_observation_logscale}(a) 
shows the pre-maximum halt on day 1.0. 
This brightness is comparable to the left edge of
the free-free emission model $V$ light curve
(thin black line) calculated from  
the steady-state wind solutions \citep{kat94h}.
Here, we adopt the $1.2 ~M_\sun$ WD model for the chemical composition
of the hydrogen-rich envelope, Ne nova 2
\citep[Ne2,][]{hac10k, hac14k, hac25kv392per}.  The Ne2 chemical
composition is listed in the figure, i.e., $X=0.55$, $Y=0.30$, $Z=0.02$,
$X_{\rm CNO}=0.10$, and $X_{\rm Ne}=0.03$ by mass.  

In case of V1674 Her, the pre-maximum halt is not clear but can be identified
at $M_V\approx -9.2$ in Figure \ref{v1674_her_optical_peak_linear}(a)
and (c).  If we take the shocked shell photosphere light curve of the 
solid magenta line both in panel (a) and (c),  they cross the black line
(free-free emission) near the pre-maximum halt phase on 0.35 day.
This suggests that a strong reverse shock arises, at least on day 0.35,
soon after the pre-maximum halt in V1674 Her.
Therefore,  we expect that a strong shock also
arises in V1500 Cyg just after the pre-maximum halt. 
Hereafter, we assume that an optically thick shocked shell is formed
on day 1.4 in V1500 Cyg.  The upward and downward arrows labeled
``shock'' in Figure \ref{v1500cyg_1674her_model_v_observation_logscale}(b)
indicate both the epochs of shock arising for V1500 Cyg and V1674 Her,
respectively.

A close look at Figure \ref{v1500_cyg_optical_peak_linear}(a)
shows the $V$ magnitude jumps up by 0.7 mag from 
$M_V= -9.32$ ($V=2.98$) on day 1.227 to $M_V=-10.0$ ($V=2.3$) on day 1.451.
This phase is almost coincident with the shock arising phase on day 0.32
of V1674 Her or upward and downward arrows labeled ``shock'' as indicated
in Figure \ref{v1500cyg_1674her_model_v_observation_logscale}(b).

We started our luminosity calculation of the shocked shell on day 1.4
at the photospheric radius of $R_{\rm ph,sh}=290 ~R_\sun$
with its expansion velocity of 1700 km s$^{-1}$ which was observed
on day 2.36 \citep{bol76g, feh76a}.
Figure \ref{v1500_cyg_optical_peak_linear}(a) shows our best fit model 
of V1500 Cyg among various trial ones of $R_{\rm ph,sh}$ on day 1.4
like in Figure \ref{v1674_her_optical_peak_linear} for V1674 Her,
i.e., the evolution of the absolute $V$ magnitude of the optically
thick shocked shell (blue line labeled $M_V$ in 
Figure \ref{v1500_cyg_optical_peak_linear}(a)). 
The corresponding photospheric temperature $T_{\rm ph,sh}$ (green line)
and radius $R_{\rm ph,sh}$ (red line) are plotted
in Figure \ref{v1500_cyg_optical_peak_linear}(b).
Here, $R_{\rm ph,sh}$ and $T_{\rm ph,sh}$ are the photospheric radius
and temperature of the shocked shell as shown in Figure 
\ref{wind_shock_config}(d).

Adopting the shell mass of $M_{\rm shell}\sim 3\times 10^{-6} ~M_\sun$,
we calculate the optical depth (Equation (\ref{potical_depth_shell}))
of the shocked shell, which decreases with time from 
$\tau_{\rm shell} \sim 3$ (day 1.4) to $\tau_{\rm shell} \sim 0.3$ (day 4.4)
as shown in Figure \ref{v1500_cyg_optical_peak_linear}(a).
The Brackett-$\gamma$ line is seen in absorption on day 2.8 while it is
seen in emission on day 3.8 \citep{enn77}, indicating that 
the shocked shell became optically thin ($\tau_{\rm shell} \lesssim 1$) on day 
$\sim$3.  This change of the optical depth $\tau_{\rm shell}$ is consistent
with the magenta line in Figure \ref{v1500_cyg_optical_peak_linear}(a). 

We regard that the shocked shell becomes sufficiently optically thin on day 
4.5--5.0, where the $V$ luminosity of the shocked shell photosphere
becomes fainter than that of free-free emission as shown in
Figure \ref{v1500cyg_1674her_model_v_observation_logscale}(a).
The radius of the shocked shell (or the photosphere of recombination
front) reached $R_{\rm ph,sh}\sim 950 ~R_\sun$ on day 4.6.

In Figure \ref{v1500_cyg_optical_peak_linear}(b), we added 
the evolution of blackbody temperature (open black circles)
of the shocked shell estimated by \citet{gal76}.
Although their obtained temperatures are slightly lower than our 
model temperature evolution (green line), their evolutionary trend is
broadly consistent with our model line.
Their spectral energy distribution ($\lambda F_\lambda$ against $\lambda$) 
during the optically thick shocked shell phase (from day 1 to day 4)
does not seem to reach maximum in their Figure 1, which could prevent
accurate determination of the blackbody temperature, where 
$F_\lambda$ is the flux at the wavelength $\lambda$.
Possibly the peak $(\lambda F_\lambda)_{\rm max}$ is located 
outside the wavelength range of their Figure 1, i.e.,
at $\lambda \lesssim 5500$\AA\  of the $V$ band, 
or at $\lambda \lesssim 4400$\AA\  of the $B$ band.
This is the reason why their blackbody temperature is lower
than our best fit temperature evolution as shown in Figure
\ref{v1500_cyg_optical_peak_linear}(b).

Although $M_{V, \rm max}\approx -10.2$ at the peak of V1500 Cyg,
\citet{gal76} expected $M_{V,\rm max}\approx -9.2$
from the MMRD relation \citep{ros65}.
This is another reason for their lower blackbody temperatures. 
If we adopt the brighter value of $-10.2$ mag instead of $-9.2$ mag
for the same distance, extinction, and photospheric radius, the blackbody
temperature should increase to 1.25 times higher than their estimates
because of $L_{\rm ph}=4\pi R_{\rm ph}^2 \sigma T_{\rm ph}^4$.  In Figure
\ref{v1500_cyg_optical_peak_linear}(b), we plot the temperature on day 1.8 by
1.25 times increase (open magenta circle pointed out by a black arrow),
which is just on our best fit evolution line (green line).

We also estimate the blackbody temperature from the intrinsic $(B-V)_0$ 
color by the relation of $(B-V)_0= (B-V) - E(B-V)= (B-V)-0.43$ and
plot them (from day 1.4 to day 3) by open green circles
in Figure \ref{v1500_cyg_optical_peak_linear}(b).
Here, we have adopted the early phase data of $B-V$ from IAUC
2826, 2828, and 2830.  These blackbody temperature data
follow well our model line (green line), supporting our optically thick
shocked shell model.

Figure \ref{v1500_cyg_optical_peak_linear}(b) also shows 
the evolution of the photospheric radii (open blue squares) 
of the shocked shell, calculated from the angular diameters $\theta$ 
taken from Figure 3 of \citet{gal76} in units of milliarcsecond ($0\farcs001$).
Here, we fit their milliarcsecond evolution of the photospheric diameter
$\theta$ with our expansion velocity of 1700 km s$^{-1}$ at the distance
of 1.5 kpc. Here, we use the relation of
\begin{eqnarray}
R_{\rm ph,sh} &=& 211~R_\sun \left( {{t}\over{{\rm 1~day}}} \right) 
\left( {{v_{\rm shell}} \over {1700 {\rm ~km~s}^{-1}}} \right) \cr\cr\cr
& \approx & {{\rm au} \over {1.5}}  
\left( {2000 {\rm ~km~s}^{-1}} \over {1700 {\rm ~km~s}^{-1}} \right)
\left( {{d} \over {1.5 {\rm ~kpc}}} \right)^{-1} \cr\cr\cr
 & &  \times {0\farcs001} \left[ \left( {{t}\over{{\rm 1~day}}} \right) 
\left( {{v_{\rm shell}} \over {2000{\rm ~km~s}^{-1}}} \right) \right] \cr\cr
& \approx & 0.78 ~\theta~{\rm au},
\label{radius_expansion_v1500_cyg}
\end{eqnarray}
where $t$ is the day after the outburst, au is the astronomical unit,  
$v_{\rm shell}$ is the expansion velocity of the shocked shell,
and $d$ is the distance to V1500 Cyg.  
Note that $211 ~R_\sun \approx 216 ~R_\sun = 1$ au.
\citet{gal76} assumed that $d=1.5$ kpc and
$v_{\rm shell}= 2000$ km s$^{-1}$, and their $\theta$ (angular
diameter) of their Figure 3 in units of milliarcsecond
represents the value in the bracket of 
Equation (\ref{radius_expansion_v1500_cyg}), i.e.,
$\left[ \left( {{t}/{{\rm 1~day}}} \right) 
\left( {{v_{\rm shell}}/{2000{\rm ~km~s}^{-1}}} \right) \right]$.

\citet{gal76}'s photospheric radii (open blue squares)
follow well our model line (red line)
in Figure \ref{v1500_cyg_optical_peak_linear}(b).
Thus, we conclude that the shocked shell photosphere dominates
the $V$ luminosity near optical peak between day 1.4 and 4.6.
We see the recombination front of the shocked shell
so that the optical spectra show a simple P-Cygni profile
with the absorption velocity of $\sim 1700$ km s$^{-1}$.

We plot this best fit model (blue line) in Figure 
\ref{v1500cyg_1674her_model_v_observation_logscale}(a) and (b).
This model $V$ light curve (blue line) follows well the observation.
The photospheric temperature decreases linearly from $T_{\rm ph,sh}=10000$ K
on day 1.4 to $T_{\rm ph,sh}=4500$ K on day 4.6, as shown
in Figure \ref{v1500_cyg_optical_peak_linear}(b).
This temperature decreasing trend is very consistent with the decreasing
trend of the observed blackbody temperatures estimated by \citet{gal76}
or calculated from the observed $B-V$ colors, 
as plotted in Figure \ref{v1500_cyg_optical_peak_linear}(b).
Thus, we reproduce the $V$ light curve of superbright nova V1500 Cyg
around the optical peak by our optically thick shocked shell model.

\section{Discussion}
\label{sec_discussion}

\subsection{Spectra from the optically thick shocked shell}
\label{shocked_shell_thick_model}

An optical spectrum of V1674 Her acquired on day 0.747 near optical maximum
\citep[$=$UT 2021 June 12.923;][]{hab24} reminds us optical spectra
of SNe IIP in the optical plateau phase.
This spectroscopic and photometric feature of SNe IIP is attributed
to the recombination front of hydrogen in shock-heated expanding SN-ejecta
\citep[see, e.g, ][ for SNe IIP spectra]{dub25}. 
\citet{hab24} also noted that, at this stage, a significant portion of the
line-forming region was optically thick. 

In our $1.35 ~M_\sun$ WD model, a strong shock arises soon after the 
maximum expansion of the WD photosphere on day 0.32, as already explained
in Section \ref{shock_formation}.
The shock heated temperature is estimated to be $\sim 4$ keV 
\citep[$\approx 4\times 10^7$ K;][]{kat25hs}
and the shocked shell is expanding at $v_{\rm shock}\sim 3000$ km s$^{-1}$
\citep{kat25hs}.  If the shocked shell is optically thick, the recombination
front lies slightly outside the shock as illustrated in Figure
\ref{wind_shock_config}(d).  Here, the recombination front corresponds to
the photosphere of the shocked shell.
It should be noted that the WD photosphere is located inside of the optically
thick shocked shell as illustrated in Figure \ref{wind_shock_config}(b).
As a result, we observe a simple P-Cygni profile of H$\alpha$ (or H$\beta$)
line with the shell velocity of $\sim 3000$ km s$^{-1}$
\citep[see Figure 2(a) of ][]{hab24}.

\citet{ayd21sc} reported three spectra on day 0.75, 0.83, and 1.74. 
The first and second spectra belong to the optically thick shocked shell
phase while the third one does to the optically thin shocked shell phase.
Their first spectra show simple P-Cygni profiles of Balmer, \ion{He}{1}, and
\ion{Fe}{2}.  The absorption troughs of the P-Cygni profiles are
at blue-shifted velocities between 3000 and 3500 km s$^{-1}$,
which is consistent with the results on day 0.66--0.67 
reported in \citet{mun21vd}.
The third spectra on day 1.74 show significant changes
and is dominated by broad emission lines of the same species
with shallow blue-shifted absorptions. The FWZI of the Balmer lines
is $> 11000$ km s$^{-1}$ and the troughs of the absorption features are
at blue-shifted velocities of around 5000 km s$^{-1}$.
These high velocity components come from the inner winds inside of the shocked
shell as illustrated in Figure \ref{wind_shock_config}(c).
These broad lines correspond to ``diffuse enhanced'' absorption/emission
line systems proposed by \citet{mcl42}. 
A similar transition of absorption line features were noted by \citet{hab24}
from day 0.747 (optically thick shocked shell)
to day 1.767 (optically thin shocked shell). 
 
\subsection{Optically thin shocked shell spectra and hard X-ray emission}
\label{shock_optically_thin}

The shocked shell of V1674 Her expands
 and becomes optically thin from day $\sim$1.1--1.2.
We are able to observe the high velocity component ($\sim$5000 km s$^{-1}$)
of the inner winds inside of the shocked shell
and the higher temperature nova (WD) photosphere,
as illustrated in Figure \ref{wind_shock_config}(c),
and as described in the previous subsection,
Section \ref{shocked_shell_thick_model}.

Swift observed V1674 Her every 1--2 days, from 1.31 days after the outburst
\citep{dra21}.  The Swift/X-ray telescope (XRT) detected the X-ray
from V1674 Her on day 1.31 \citep[see Figures 
\ref{v1674_her_v_x_observation_only_logscale} and
\ref{v1674_her_v_x_observation_only_logscale_no2_final}; also Figure 1 
of ][]{dra21}.
The hardness ratio of (1--10 keV)/(0.3--1 keV)
is as high as 100, so that they are hard X-rays and emitted from 
shock heated optically thin thermal plasma.
This is broadly consistent with our result
that the shocked shell became optically thin after day $\sim 1.2$.

\subsection{Distance and Reddening}
\label{sec_distance}

The distance to V1674 Her is not well constrained.
Various authors have presented various values from $\sim 2$ to $\sim 6$ kpc
\citep{bai21rf, woo21, qui24, schaefer22, sok23, hab24}.
A negative Gaia eDR3 parallax is given by \citet{bai21rf} to be
$\varpi= (-0\farcs 93628335 \pm 0\farcs 6273195)\times 10^{-3}$,
indicates a rather long distance (we suppose $d\gtrsim 5$ kpc).

\citet{schaefer22} gave a rather small distance of
$d=$3216 (2472--5329) pc based on the Gaia eDR3 parallax while
\citet{bai21rf} listed a different distance of
$d=$6000.153 (3242.231--9802.509) pc from the same negative parallax.
These authors assumed different priors for Bayesian inference
(prior probabilities of Bayesian statistics), in other words,
they assumed different 3D distributions of stars/novae in our Galaxy
as a prior.  
This simply means that the assumption is the result for a negative
parallax.  Therefore, we are not able to accurately constrain
the distance to V1674 Her only with the Gaia eDR3 parallax. 

\citet{sok23} obtained $d=6.3^{+3.8}_{-2.4}$ kpc from the statistical
relation between the luminosity and the orbital period for IPs
\citep{war87, muk23p}, that is, $M_V= 4.8\pm 1$ in quiescence for the orbital
period of V1674 Her, $P_{\rm orb}=3.67$ hr \citep{pat22}. 
They criticized the other shorter distances \citep{schaefer22,
woo21, dra21} and adopted $d=6.3$ kpc.

\citet{kat25hs} estimated the distance to V1674 Her
to be $d=8.9\pm 1$ kpc using the $V$ band distance modulus
$(m-M)_V= 16.3\pm 0.2$ and the distance-reddening relation of
Galactic 3D extinction map given by \citet{gre19}. 
Here, we have already checked the distance modulus of $(m-M)_V= 16.3\pm 0.2$
toward V1674 Her by the time-stretching method in Section
\ref{sec_timestretching}.
From the crossing point of Equation (\ref{distance_reddening_law}) 
with $(m-M)_V= 16.3$ and \citet{gre19}'s distance-reddening relation,
\citet{kat25hs} obtained $d=8.9\pm 1$ kpc and $E(B-V)= 0.5 \pm 0.05$
\citep[see Figure 10 of ][]{kat25hs}.
The reddening of $E(B-V)= 0.5\pm 0.05$ is consistent with the reddening of
$E(B-V)=0.55$ estimated from the interstellar absorption feature of
\ion{K}{1} 7699 by \citet{mun21vd},
and is also supported by \citet{schlaf11f}'s 2D 
Galactic reddening map of $E(B-V)= 0.4985\pm 0.0191$ toward V1674 Her.

In the present paper, we adopt $(m-M)_V= 16.3$, $d=8.9$ kpc,
and $E(B-V)= 0.5$.  The distance of $d=8.9\pm 1$ kpc is broadly
consistent with \citet{bai21rf}'s 
$d = 6.0^{+3.8}_{-2.8}$ kpc and \citet{sok23}'s $d=6.3^{+3.8}_{-2.4}$ kpc.
The distance modulus of $(m-M)_V= 16.3$ is also supported by
an excellent fit with the free-free emission model light curve of
a 1.35 $M_\sun$ WD outburst model ($\dot{M}_{\rm acc}=1\times 10^{-11}
~M_\sun$ yr$^{-1}$) except for around the optical peak
as shown in Figure \ref{v1674_her_v_x_observation_only_logscale}.

\section{Conclusions}
\label{sec_conclusion}

V1674 Her is the fastest ($t_2\sim 0.9$ day) and brightest
($M_{V,\rm max}\sim -10.2$) nova in our Galaxy, which belongs to
the class of superbright novae defined by \citet{del91}.
We elucidate the origin of this superbright nova. 
Our results are summarized as follows: \\
\begin{enumerate}
\item Our $1.35 ~M_\sun$ WD model with the mass-accretion rate of 
$1\times 10^{-11} ~M_\sun$ yr$^{-1}$ shows that the WD envelope expands
to blow strong winds and reaches maximum expansion of
the photosphere 0.32 days after the outburst.
A strong reverse shock arises on day 0.32 and
GeV gamma-rays are emitted from this shocked layer.
This is consistent with the gamma-ray detection on day 0.39.
\item We elucidate that the shocked shell is optically thick
around the $V$ peak, during which we observe the photosphere
of the recombination front of hydrogen in the shocked shell
that is located far outside the nova (WD) photosphere. 
\item We have calculated the $V$ light curve from the shocked shell
photosphere.  Its $V$ brightness increases with time as the shell
expands with velocities between $\sim 2600$ km s$^{-1}$ and
$\sim 4000$ km s$^{-1}$.  Theoretically,
the optical $V$ brightness reaches maximum on day $\sim$0.5--0.7.
After the maximum, the shocked shell becomes optically thin
and the shell brightness drops.  This kind of drops in optical are
also observed in the recombination front of hydrogen
as seen in shock-heated expanding SNe IIP ejecta.
Observationally, the $V$ magnitude reaches maximum on day $\sim$0.7--0.8
because of no visual or $V$ observation between day 0.5 and 0.7.
\item 
Our model $V$ light curves reasonably reproduce the shape of the $V$ peak.
The absolute $V$ brightness attains its maximum of
$M_V\sim -10.2$ mag when the photosphere of the shocked shell expands
to $R_{\rm ph,sh}\sim 300 ~R_\sun$ because 3500 km s$^{-1}~\times$ 0.7 days
$=$ 300 $~R_\sun$. 
This confirms our expectation that the peak magnitude of 
superbright nova V1674 Her is owing to the contribution of
an optically thick shocked shell.
\item 
GeV gamma-rays were detected on day $\sim$0.4 
clearly before the optical peak on day 0.5--0.7 from the outburst.  
This is the first case that the GeV gamma-ray peak substantially
precedes the optical maximum.
The chronological order of shock formation,
emergence of GeV gamma-rays, and optical maximum are naturally explained 
because the observed optical peak is not the peak of free-free emission 
of nova winds, but the peak of the shocked shell photosphere that is
located far outside the nova (WD) photosphere.
\item The shocked shell expands and becomes optically thin on day
$\sim$1.1--1.2.  After that, we are able to observe the high-velocity
components $\sim$5000 km s$^{-1}$ of the inner winds inside of
the shocked shell and the higher temperature nova (WD) photosphere.
A transition of the spectrum from simple P-Cygni absorption
($\sim$3000 km s$^{-1}$) troughs to broad emission lines of the same species
with shallow blue-shifted ($\gtrsim$5000 km s$^{-1}$) absorptions
occurred after day $\sim$1.1--1.2. 
\item Hard X-rays from optically thin plasma were detected with the 
Swift/XRT on day 1.31, which is consistent with the epoch that
the shocked shell becomes optically thin. 
\item V1500 Cyg is a prototype of superbright novae defined by
\citet{del91} in our Galaxy.
The evolution of spectral energy distribution clearly shows that
the nova spectrum changed from blackbody to free-free emission 
on day 4--5.  This is the same transition as that of V1674 Her,
from optically thick, shocked shell to optically thin shell
on day 1.2.  Thus, we conclude that the optically thick, shocked shell plays 
an essential role in the peak $V$ brightness of superbright novae.
\item 
We obtain $(m-M)_{V, \rm V1674~Her}=16.3$ for V1674 Her with the 
time-stretching method. i.e., substituting V1674 Her as the target
and V1500 Cyg as the template novae into 
Equation (\ref{distance_modulus_formula}) together with the $V$ band
distance modulus of $(m-M)_{V, \rm V1500~Cyg}=12.3$ after \citet{hac14k}.
This result is consistent with that obtained by \citet{kat25hs}
with the same time-stretching method but against other three template novae,
LV Vul, V339 Del, and KT Eri.
\end{enumerate}

\begin{acknowledgments}
We acknowledge with thanks the variable star observations (V1674 Her)
from the AAVSO International Database contributed by observers worldwide
and used in this research.
We are also grateful to the anonymous referee for useful comments 
that improved the manuscript. 
\end{acknowledgments}

\vspace{5mm}
\facilities{Swift(XRT), AAVSO}



\end{document}